%% file: arXiv.tex
\documentclass[journal,transmag]{IEEEtran}
\IEEEoverridecommandlockouts
\usepackage{cite}
\usepackage{amsmath,amsthm,amssymb,amsfonts}
\usepackage{algorithm}
\usepackage{algorithmicx}
\usepackage[noend]{algpseudocode}
\usepackage{graphicx}

\usepackage{textcomp}
\usepackage{xcolor}
\usepackage{pifont}
\usepackage{forest}
\usepackage{verbatim}
\usepackage{adjustbox}
\usepackage{pdfpages}
\usepackage{balance}  
\usepackage{paralist}
\usepackage{subfigure}
\usepackage{caption}
\usepackage{url}

\algnewcommand\algorithmicinput{\textbf{Input: }}
\algnewcommand\INPUT{\item[\algorithmicinput]}
\algnewcommand\algorithmicoutput{\textbf{Output:}}
\algnewcommand\OUTPUT{\item[\algorithmicoutput]}
\algnewcommand\algorithmicsubroutine{\textbf{Subroutine: }}
\algnewcommand\Subroutine{\item[\algorithmicsubroutine]}
\newtheorem{definition}{Definition}
\newtheorem{lemma}{Lemma}

\makeatletter

\newcommand{\Rmnum}[1]{\expandafter\@slowromancap\romannumeral #1@}
\makeatother
\newcommand{\basis}{\newline \noindent {\bf Basis:} }
\newcommand{\ih}{\newline \noindent {\bf Induction hypothesis:}  Assume }
\newcommand{\is}{\newline \noindent {\bf Induction:} }

\newcommand{\para}[1]{\smallskip\noindent\textbf{#1}}
\usepackage[font=bf]{caption}

\def\BibTeX{{\rm B\kern-.05em{\sc i\kern-.025em b}\kern-.08em
    T\kern-.1667em\lower.7ex\hbox{E}\kern-.125emX}}
\begin{document}

\title{Efficient Error-tolerant Search on Knowledge Graphs}

\author{\IEEEauthorblockN{Zhaoyang Shao}
\IEEEauthorblockA{\textit{University of Alberta} \\
zhaoyang@ualberta.ca}
\and
\IEEEauthorblockN{Davood Rafiei}
\IEEEauthorblockA{\textit{University of Alberta} \\
drafiei@ualberta.ca}
\and
\IEEEauthorblockN{Matteo Lissandrini}
\IEEEauthorblockA{\textit{Aalborg University} \\
matteo@cs.aau.dk}
\and
\IEEEauthorblockN{Themis Palpanas}
\IEEEauthorblockA{\textit{Univ. of Paris \& French Univ. Inst. (IUF)} \\
themis@mi.parisdescartes.fr}
}

\IEEEtitleabstractindextext{%
\begin{abstract}
\input{sections/00_abstract.tex}
\end{abstract}

}
\maketitle

\newcommand*{\Archive} 

\IEEEdisplaynontitleabstractindextext
\IEEEpeerreviewmaketitle

\input{sections/01_intro.tex}
\input{sections/02_motivating.tex}
\input{sections/03_problem.tex}
\input{sections/04_approach.tex}
\input{sections/05_cost.tex}
\input{sections/05p_extensions.tex}

\input{sections/06_experiments.tex}

\input{sections/07_related.tex}
\input{sections/08_conclusions.tex}
\input{sections/09_acks.tex}

\bibliographystyle{IEEEtran}
{
\bibliography{ref}  %
}

\input{sections/99_appendix}


\end{document}

%% file: sections/00_abstract.tex
Edge-labeled graphs are widely used to describe relationships between entities in a knowledge graph.
Given a query subgraph that represents an example of what the user is searching for, we study the problem of efficiently searching for similar subgraphs in a large knowledge graph, where the similarity is defined in terms of the well-known graph edit distance. We call these queries {\em error-tolerant exemplar queries} since matches are allowed despite small variations in the graph structure and the edge labels.
We propose two efficient exact algorithms, based on a filtering-and-verification framework, for finding subgraphs in a large data graph that are isomorphic to a query graph under edit operations. Our filtering scheme, which uses the neighborhood structure around a node and the presence (or absence) of paths, significantly reduces the number of candidates that need to be processed by the verification stage. 
We develop a cost model that characterizes the relationships between different variables (e.g., node degrees and edit distance threshold) and reveals some of the settings that affect performance and the conditions under which our algorithms are expected to perform well.
Our experimental evaluation further reveals the effectiveness of our 
filtering schemes and queries, the efficiency of our algorithms, as well as the reliability and accuracy of our cost models, on real datasets and query workloads.

%% file: sections/01_intro.tex
\section{Introduction}
\label{sec:intro}
\IEEEPARstart{K}{nowledge} graphs have become a cornerstone of many applications that rely on connecting facts about heterogeneous entities gathered from different sources.
However, users querying a knowledge graphs often are not familiar with the structure of the graph and the vocabulary adopted, hence exploratory search is crucial even for those who are skilled in standard query languages such as SQL and SPARQL.

\begin{figure*}[tb]
\centering
\hspace*{-0.4cm}
\includegraphics[width=1.05\textwidth]{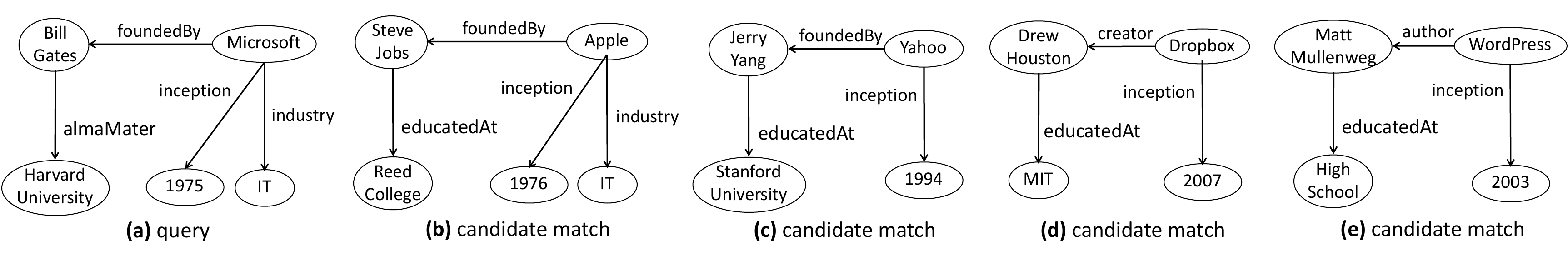}
\caption{(a) An ETEQ query, and (b), (c), (d) and (e) matches at different edit distances}
\label{fig:eteq-query-ex}
\end{figure*}

One class of exploratory search is exemplar queries~\cite{mottin2016exemplar}, where users provide an example of what they are searching for.
For example,  a user searching for information about founders of technology companies may provide, as an example, the relationships between ``Bill Gates,'' ``Microsoft,'' ``Harvard University'' and ``1975'', as  shown in Figure~\ref{fig:eteq-query-ex}-a.
These relationships, expressed in a graph query, can retrieve other entities with the same, but not similar relationships among them.
For example, such a query on Wikidata will not retrieve any matches since the predicate \emph{almaMater} does not exist.
Suppose the user finds out that the corresponding predicate in Wikidata is \emph{educatedAt} and changes the query accordingly.
The revised query will retrieve tuples such as (``Steve Jobs'', ``Apple'', ``Reed College'', ``1976''), but will still miss WordPress, which has an \emph{author} but not \emph{foundedBy}, and Dropbox, which has a \emph{creator} but not \emph{foundedBy}.
The revised query will also miss Yahoo, which has no industry predicate. 

Enumerating all these variations at querying time is a tedious and time-consuming task.
This paper studies the problem of efficiently supporting exploratory searches on knowledge graphs under such variations.
In practice, this requires finding subgraphs that are composed of both exact and similar relationships and those relationships form similar structures to the user query, while ignoring matches on the entity labels.

In general, searching for similar, rather than exact, matches of a query is more desirable when the user is not aware of the graph structure and/or the label data is noisy, or when inconsistencies are allowed.
For example, in computational biology, the data can be highly noisy because of possible errors in data collection, different thresholds used in experiments, as well as the difficulty in cleaning the data.
Despite the noise, searching for similar biological structures may enable a biologist to learn more about a new organism~\cite{dost2008qnet}.
In molecular chemistry, identifying similar molecular structures of a target molecule may enable a chemist to design new molecular structures~\cite{balaban1985applications}.
In social network analysis, searching for similar subgraphs may help to identify communities and to predict the network dynamics~\cite{spertus2005evaluating}.
In all aforementioned scenarios, one needs to identify the existing subgraphs in a data graph that are similar to a query graph.

Despite the large body of work on subgraph search,
many of the techniques in the literature cannot address the aforementioned scenarios, where knowledge graphs are queried using examples.
For example, a number of works have tackled this problem for graph databases~\cite{tian2008tale,mongiovi2010sigma,wang2012efficiently,zeng2009comparing} where special indexes are computed, but these approaches are either not applicable, or not efficient when there is a large single data graph.
The most similar work to ours is SAPPER~\cite{zhang2010sapper}, but this work only allows edge removal and not edge renaming (nor the equivalent deletion followed by insertion).
Moreover our experiments show worse performance for SAPPER in terms of running time.
Other existing approaches for approximate graph search do not tackle exact measures like edit distance, and are approximate in the sense that they do not guarantee that all valid answers will be retrieved~\cite{khan2013nema,dutta2017neighbor}.
There is also work on query rewriting for RDF graphs, which does not address the problem of fast retrievals, and requires external knowledge about which reformulations are permitted and in which form~\cite{zheng2016semantic,huang2012approximating,elbassuoni2011query}.

\para{Error-Tolerant Exemplar Queries.}
In this paper, we propose a framework that overcomes the problems mentioned above, through the use of graph edit distance operations in \emph{exemplar queries}~\cite{mottin2016exemplar}, which allows a principled and well defined notion of similarity.
Nevertheless, introducing edit operations in exemplar queries can significantly expand the search space. For example, limiting the edit operations to edge substitution and with $L$ possible edge labels, a naive solution is to run an exemplar query for every edit. For a query graph with $|E_q|$ edges, the number of such exemplar queries is $O(L |E_q|)$ for edit distance threshold $1$, and
$O(L^t {|E_q| \choose t})$ for edit distance threshold $t$, which is prohibitively expensive.
Therefore, novel techniques are necessary in order to provide efficient and scalable solutions.
We call our queries \emph{error-tolerant exemplar queries (ETEQ)} since
mismatching edge labels are allowed under edit operations.

Given that ETEQ generalizes exemplar queries, the queries in ETEQ are applicable in many domains, where the user does not have a concrete formulation of what is being searched, but can identify an element from the desired result set.

There are a number of challenges in efficiently evaluating ETEQ.
First, for a query with $|E_q|$ edges, we will need $O(|E_q|)$ joins to put together the results of edge matches.
This becomes a computationally intensive process
for large values of $|E_q|$.
Second, allowing edit operations further increases the size of the search space, as well as the space overhead for the intermediate results.
We address these challenges by (1) proposing efficient indexes and sketches for filtering candidates; and (2) developing novel, accurate estimates for query selectivity and cost.
Thus, we describe two new algorithms for efficiently evaluating ETEQ: these algorithms explore the overlap among query transformations under different edit operations, and can effectively reduce the search space and minimize the overall cost.

\para{Contributions.}
The contributions we make in this work can be summarized as follows:
\begin{itemize}

\item
We propose exemplar queries with edit distance operations for error-tolerant exploratory searches on knowledge graphs.

\item
We present two efficient algorithms for ETEQ based on a filtering-and-verification framework, and study efficient pruning strategies that use the neighborhood structure and the paths to filter unqualified results.

\item
We develop a novel cost model that allow us compare the expected performance of our algorithms across different queries, before actually running them.

\item
We analyze the performance of our algorithms using the proposed cost model, and study the conditions under which each algorithm is expected to outperform well or not so well.

\item
We perform a thorough experimental evaluation, using real data and query workloads, of the effectiveness of our filtering schemes, the performance and the scalability of our algorithms, as well as the reliability of our cost model.
The results demonstrate the efficiency and effectiveness of the proposed approach.
\end{itemize}

\noindent{\bf Paper outline.}
The rest of this paper is organized as follows.
Section~\ref{sec:example} presents a motivating example, and Section~\ref{sec:def} formally defines the problem.
We describe our approach in Section~\ref{sec:approach}, and provide a cost analysis study in Section~\ref{sec:costanalysis}.
Further refinement of ETEQ queries are discussed in Section~\ref{sec:extensions}.
We present the experimental evaluation in Section~\ref{sec:experiments} and the related work in Section~\ref{sec:related}.
Finally, we conclude in Section~\ref{sec:conclusions}.

%% file: sections/02_motivating.tex
\section{Motivating Example}
\label{sec:example}





Suppose we want to find the founders of technology companies in a knowledge graph, and may provide, as an example, the relationships among ``Bill Gates'', ``Microsoft'', ``Harvard University'', and ``1975'', as  depicted in Figure~\ref{fig:eteq-query-ex}(a). 
The same figure shows some of the answers at edit distance 1 in Figure~\ref{fig:eteq-query-ex}(b), edit distance 2 in Figure~\ref{fig:eteq-query-ex}(c), and edit distance 3 in both Figure~\ref{fig:eteq-query-ex}(d) and Figure~\ref{fig:eteq-query-ex}(e).

Note that none of the possible answers shown in Figure~\ref{fig:eteq-query-ex} are graph isomorphic to the query.
In an exploratory search using graph isomorphism (which is equivalent to setting the edit distance threshold to zero in our case), a user issuing the query in Figure~\ref{fig:eteq-query-ex}(a) would obtain an empty result set. Through some trial and error, the user may notice that none of the possible matches have the predicate ``almaMater'', and that the predicate should be changed to ``educatedAt'' to find more matches. Changing the query by replacing ``almaMater'' with ``educatedAt'',  will reduce the distances of the matches in Figure~\ref{fig:eteq-query-ex}(b), (c), (d), and (e) to 0, 1, 2 and 2, respectively.
This means that the user will retrieve only Figure~\ref{fig:eteq-query-ex}(b) as an answer, and will require further reformulations to obtain more results.

This example highlights the need for an error tolerant search algorithm that allows both for edge removal and edge renaming to enable exploratory searches.
%

%% file: sections/03_problem.tex
\section{Problem Definition}
\label{sec:def}
As we study error-tolerant exemplar queries on knowledge graphs, we provide a few definitions before formally introducing the problem.
Knowledge graphs are usually treated as labeled directed graphs where nodes represent entities and edges represent their relationships.
\begin{definition}{(Knowledge Graph)}
A knowledge graph $G$$=$$\left \langle V, E, L\right\rangle$ is a directed labeled graph, where $V$ denotes a set of nodes, $E \subseteq V^2$ is a set of edges, and $L$ is a labeling function that maps each node and each edge to a label.
\end{definition}

Unless explicitly stated otherwise, the terms \emph{graph}, \emph{knowledge graph} and \emph{edge-labeled graph} are used interchangeably in this paper.

\begin{definition}{(Edge-preserving Isomorphism)}
\label{def:edgeiso}
A graph $G$ is edge-preserving isomorphic to a graph $G'$, denoted as $G\simeq G'$, if there is a bijective function $\mu$ from the nodes of $G$ to the nodes of $G'$ such that for every edge $n_1 \xrightarrow{l} n_{2}$ in $G$, the edge $\mu(n_1) \xrightarrow{l} \mu(n_2)$ in $G'$.
\end{definition}

\begin{definition}{(Edge-preserving Edit Distance)}
The edit distance between two non-isomorphic graphs $G$ and $G'$ is the minimum number of edit operations that makes $G \simeq G'$.
\end{definition}

\begin{definition}{(Error-tolerant Exemplar Query)}
\label{def:eteqQ}
An error-tolerant exemplar query is a pair $(Q,t)$ where $Q$ is a connected graph and $t \in R$ is a threshold. The answer to query $(Q,t)$ on a knowledge graph $D$ is the set of all subgraphs $S$ in
$D$  such that $S$ becomes edge-isomorphic to $Q$ after applying some edit operations to $Q$, $S$ or both, and the cost of those operations does not exceed the threshold $t$.
\end{definition}

Edit operations generally include insertion, deletion and substitution of edge or node labels, and these operations may be applied to both query and data graphs.
Without loss of generality, we limit the edit operations only to the queries.
Note that not all edit operations are applicable in an edge-preserving isomorphism, where the relationships (edges), and not the entities (nodes), are considered in the match.
In particular, label substitutions are limited to edge labels, and inserting an edge to the query graph may be ignored (this would actually restrict a query, since the new edge will be an additional constraint that the candidate answers would have to satisfy, while our goal is to relax the query).
This reduces the edit operations to edge deletion (which simplifies and relaxes a query) and edge label substitution (which enables matching with synonyms).

Observe also that in general, each edit operation may have a different cost.
For example, substituting a label may be less costly when the two labels are synonyms.
To simplify the presentation, we assume all edit operations have the same cost, and may sometime refer to the edit threshold $t$ as the number of edit operations that are allowed.
We will refer to error-tolerant exemplar queries simply as queries.

Despite the aforementioned assumptions, our approach can be further generalized, removing some of the constraints above.
We discuss some of these extensions in Section~\ref{sec:extensions}.


\para{Problem Statement:}
Given an ETEQ in the form of a query graph $q$ and an edit distance threshold $t$, we aim to efficiently retrieve all relevant answers in a data graph that are edge-preserving isomorphic to $q$ with edit cost at most $t$.

%% file: sections/04_approach.tex
\section{Proposed Approach}
\label{sec:approach}
\subsection{The EXED Algorithm}
Given a data graph $G\,=\,(V,\,E)$, a query $Q$ and the edit distance threshold $t$,
a naive approach to find subgraphs that are within edit distance $t$ of the query $q$ is to
compare the query with every subgraph in the data graph $G$.
Our basic algorithm for exemplar queries with an edit distance constraint (referred to as EXED)
chooses a set $N_q$ of the query nodes with $|N_q|=k$ as possible starting nodes. We refer to $N_q$ as a {\it k-subset} of the query nodes and discuss later in this section how the nodes in $N_q$ can be selected and what values of $k$ will guarantee the correctness.  
The algorithm treats each node $n_q \in N_q$ as a seed
and considers all nodes of the data graph one by one as possible mappings of the node $n_q$. For each such node $n$ in $V$,  it checks if there exists a subgraph that contains $n$ and is isomorphic to the query with at most $t$ edit operations.
The algorithm (Algorithm~\ref{EXED}) starts from a query subgraph $q$ only containing $n_q$ and a data subgraph containing $n$, and maps $n_q$ to $n$. It iteratively adds edges from $Q$ and $G$ to the mapping until the resulting subgraph of $G$ is isomorphic to $Q$ with at most $t$ edit operations. All such matching subgraphs of $G$ are retrieved.

\begin {lemma}
\label{lemma:exed-correctness}
Algorithm~\ref{EXED} correctly
finds all edge-isomorphic mappings at distance $t$.
\end{lemma}
\begin{proof}
The algorithm starts with a query node $n_q$ and finds all edge-isomorphic graphs with a node mapping to $n_q$.
However, with an edit distance threshold larger than zero, an edge-isomorphic mapping may not include the edge leading to $n_q$, and such mappings may not be discovered when the search starts from $n_q$.
The number of those mismatching query graph edges (referred to as {\it query edges} for short) cannot exceed $t$, and the correctness is guaranteed by running the search $min(t+1,|V_Q|)$ times, each time starting from a different query node in $N_q$.
\end{proof} 

\begin{algorithm}[tb]
\caption{\textproc{EXED}}
\label{EXED}
{\small
\begin{algorithmic}[1]
\INPUT Data graph $G=\,\left \langle V,\,E\right\rangle$, query graph $Q=\,\left \langle V_Q,\,E_Q\right\rangle$
\INPUT Threshold $t$
\OUTPUT Set of answers $S$
\State $S \leftarrow \emptyset$; $k \leftarrow min(t+1,|V_Q|)$
\State $N_q \leftarrow k\_subset\_of(V_Q, k)$
\For{each $n_q \in N_q$}
\For{each node $n \in V$}
    \State $s\,=\,$ \Call{SearchSimilarSubgraph}{$G$, $Q$, $n_q$, $n$, $t$}
\If{$s\,\not=\emptyset$}
	\State Add $s$ to answer set $S$
\EndIf
\EndFor
\EndFor
\State \Return $S$
\end{algorithmic}
} 
\end{algorithm}

\subsection{Neighborhood-based Pruning}
In EXED, every node $n$ of the data graph is considered a possible mapping of the query node $n_q$ and as a seed to start the search for relevant answers. This is highly inefficient since only a small fraction of data nodes might be true candidates.
To reduce this search space, one has to reduce the number of unnecessary data nodes from which the search for similar subgraphs starts~\cite{khan2011neighborhood}. 
Let's introduce a notion of neighborhood before presenting our approach to prune the search space.

\begin{figure}[tb]
\centering
\input{figures/forest-falsep.tex}
\caption{Query graph and data graph}
\label{fig:falsep}
\end{figure}
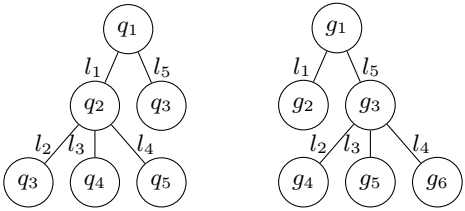

\begin{definition}{($d$-neighbor)}
\label{d-neighbor}
Let $n \in V$ be a node of the
data graph $G = \left \langle V, E\right\rangle$. The node $n_i \in V$ is a $d$-neighbor
of $n$ if there exists a path from $n$ to $n_i$ of length at
most $d$, ignoring the edge directions. The $d$-neighborhood of $n$, denoted as $N_d(n)$, is the
set of all $d$-neighbors of $n$, and the $d$-neighborhood labels of $n$, denoted as $L_d(n)$, is the
set of edge labels on paths of length at most $d$ from $n$ to its d-neighbor nodes.
For example, $N_1(q_1)=\{q_3,q_2\}$ and $L_1(q_1)=\{l_1,l_5\}$ in Fig.~\ref{fig:falsep}.
\end{definition}
 
$\textproc{NeighborhoodPruning}$ compares data nodes with query nodes using their neighborhood information, and filters out those data nodes that cannot map to a query node with an edit distance threshold $t$ based on the node neighborhoods. Let $T_{n, k, l}$ denotes those neighbor nodes of $n$ which are reachable from $n$ in a path of length $k$ and $l$ is the last label in the path, i.e.,\\
$T_{n, k, l} = \left\{ n_1 | \exists n_2 \in N_{k-1}(n) (n_1 \xrightarrow{l} n_2\cup n_1 \xleftarrow{l} n_2)\right\}$.

It can be noted that the neighborhood is insensitive to edge directions. 
Since keeping the table of neighbor nodes for every data node is expensive in term of space, we only keep the cardinality of $T_{n,k,l}$. Also, to efficiently retrieve candidate mappings of a query node, we implement an inverted index which stores a list of nodes for every label, every cardinality and every distance. In other words, the index allows us to efficiently find data nodes that have a label $l$ at their $k$-neighborhood with a certain cardinality.

Once the neighborhood tables $T_{n, k, l}$ of both data node $n$ and query node $n_q$ are computed for each label $l$ and path length $k \leq d$, then we can compare 
the cardinalities of $T_{n_q, k, l}$ and $T_{n, k, l}$ and find out the number of edge labels $l$ that are present in the $k$-neighborhood of $n_q$ but not in that of $n$. 
The edit distance between data node $n$ and query node $n_q$ for label $l$ at $k$-neighborhood can be written as
\[
dist_{k,l}(n, n_q)=
\begin{cases}
0 & \text{if } |T_{n,k,l}| \geq |T_{n_q,k,l}|\\
|T_{n_q,k,l}| - |T_{n,k,l}| &\text{otherwise}.
\end{cases}
\]
For example in Fig.~\ref{fig:falsep}, $T_{q_1, 1, l_1}=\{q_3\}$ and $T_{n_1, 1, l_1}=\{n_2\}$ and $dist_{1,l_1}(n_1, q_1)$ is zero.
Given an edit distance threshold $t$, $n$ is considered a candidate for the query node $n_q$ when the distance between the $d$-neighborhoods of the two nodes does not exceed $t$, i.e.,
$\sum_{i=1}^d\sum_{l \in L_i(n_q)} dist_{i,l}(n, n_q) \leq t$.


Note that this filtering may introduce false positives, because neighborhood-based pruning cannot identify if the labels are under the same path or have the  same direction. For example, the neighborhood-based distance between $q_1$ and $n_1$ in Fig.~\ref{fig:falsep} is $0$ whereas the actual edit distance is $2$ (i.e., renaming $(q_1,q_3)$ to $l_5$ and $(q_1,q_2)$ to $l_1$). However, it cannot introduce false negatives since the neighborhood distance can be underestimated but it cannot be overestimated. 
On the other hand, this compact representation of a neighborhood is highly effective at pruning nodes without actually visiting their neighborhood, and false positives can be removed at the verification stage. Now to compare the whole query graph, we need to run a simulation.

\begin{definition}{(Simulation)}
Let  $G_1 = \left \langle V_1, E_1\right\rangle$ and $G_2 = \left \langle V_2, E_2\right\rangle$ be two graphs. $G_2$ simulates $G_1$ if there exists  a relation $R$ such that, for every node $n_1 \in V_1$ and $n_2 \in V_2$ for which $(n_1,n_2) \in R$ and for each edge $n_1 \xrightarrow{l} n_1' \in E_1$, there exists $n_2 \xrightarrow{l} n_2' \in E_2$ such that $(n_1',n_2') \in R$.
\end{definition}

Verifying a simulation can be done efficiently since $n^\prime \in V$ is a possible mapping of $n_{q}^\prime \in V_Q$ only if in a previous comparison, node $n \in V$ is identified as a possible mapping of $n_q \in V_Q$ and there is an edge between $n$ and $n^\prime$ with label $l$ and a corresponding edge with label $l$ between $n_q$ and $n_{q}^\prime$. With this observation, we only need to examine the adjacent nodes of previously mapped data nodes rather than all data nodes to find possible mappings of a query node.

The steps of our neighborhood-based pruning and verifying a simulation are shown in Algorithm~\ref{alg:NeighbourhoodPruning}.
Similar to Algorithm~\ref{EXED}, the algorithm initially starts with a k-subset $N_q \subset V_Q$, and treats each node $s_q \in N_q$ as a seed from which the search starts. The choice of the seed can affect the performance, and we discuss next how the seed set can be selected. For each seed $s_q \in N_q$, the frontier of the search initially includes $s_q$ and the set of all data nodes are considered as candidate mappings of $s_q$. 
In each iteration of the algorithm, the neighboring nodes of query node $n_q$ are compared with a candidate nodes $\mu(n_q)$ in the data graph and either the edit distance of each node $\mu(n_q)$ is updated (if needed) or $\mu(n_q)$ is removed from the list of candidates if the edit distance exceeds the threshold. The neighborhood-based pruning is applied in Steps 8 and 9 (and before actually visiting data nodes) to prune the search space. The search for each seed $s_q$ ends when all query nodes are visited.


\begin{algorithm}[tb]
\caption{\textproc{NeighborhoodPruning}}
\label{alg:NeighbourhoodPruning}
{\small
\begin{algorithmic}[1]
\setcounter{ALG@line}{0}
\INPUT Data graph $G=\,\left \langle V,\,E\right\rangle$, query graph $Q=\,\left \langle V_Q,\,E_Q\right\rangle$
\INPUT Threshold $t$
\OUTPUT Set of candidate mappings $\mu \subset V_Q \times V $
\State $N_q \leftarrow k\_subset\_of (V_Q,min(t+1,|V_Q|)$
\For {each $s_q \in N_q$}
    \State $\text{Vis} \leftarrow \emptyset$
    \State $\text{VisFront} \leftarrow \{s_q\}$
    \State $\mu(s_q) \leftarrow (V, \vec{0})$
    \For {each $n_q \in \text{VisFront}$}
        \For {each $ \left \langle n_q, n_{q}' \right\rangle_l\in E_Q$ and $n_{q}' \not\in \text{Vis}$}
            \State Update edit distance of nodes in $\mu(n_q)$, $\mu(n_q')$.
            \State Remove nodes that exceed threshold.
        \EndFor
        \State $\text{VisFront} \leftarrow \text{VisFront} \cup \{n_q' | n_q \xleftarrow{l} n_{q}' \vee n_q \xrightarrow{l} n_{q}' \}$
        \State $\text{VisFront} \leftarrow Q \setminus \{n_q\}$
        \State $\text{Vis} \leftarrow \text{Vis} \cup \{n_q\}$
    \EndFor
\EndFor
\end{algorithmic}
} 
\end{algorithm}

The seed set $N_q$ in Step 1 is selected based on the selectivity (as defined next), where a set of $min(t+1,|V_Q|)$ nodes (see Lemma~\ref{lemma:exed-correctness}) with the least selectivity are selected.

\begin{definition}{(Selectivity)}
The selectivity of a query node $n_q$ in a data graph $G$ is the probability that an arbitrary node of $G$ maps to $n_q$. The selectivity of a label $l$, $Sel(l)$, is the probability that an arbitrary edge of $G$ is labeled $l$, and is computed as the ratio of the frequency of label $l$ to the number of edges in $G$.
\end{definition}

As the actual selectivity of a query node may be known only after finding all its mappings,
we devise a method to estimate the selectivity in advance (see Sec.~\ref{sec:costanalysis} for details).
%


\subsection{Path-based Filtering}
The neighborhood-based pruning may introduce false positives as we mentioned above. We now introduce a path-based algorithm to prune some of the false positives.

Our path-based filtering  compares data nodes with query nodes in terms of their paths and prunes those data nodes that require more than $t$ edit operations to match a query node. However, keeping every path for every node can be expensive in terms of space. For a graph with average degree $\hat{D}$, the space required for maintaining paths of length $d$ edges is $O(\hat{D}^d)$. Our solution is to use a Bloom filter, which gets close to an optimal space usage~\cite{bloom1970space}.

A Bloom filter is a space-efficient probabilistic data structure to efficiently test whether an element is a member of a set $N$. An empty Bloom filter is a bit array of $m$ bits, all set to $0$. There are $k$ different hash functions, each mapping an element to one of the $m$ array positions. To query for an element, one needs to find the $k$ array positions the element is mapped to. If any of the bits at these positions is 0, the element is definitely not in the set. If all are 1, then either the element is in the set, or the bits have by chance been set to $1$ during the insertion of other elements, resulting in a false positive. The error rate $p$ depends on $m$, $|N|$ and $k$. We set the false positive rate to  $1\%$. The optimal number of hash functions is approximately $0.7m/|N|$, and the optimal number of bits $m$ is approximately $|N|\ln{p}/\ln^2{2}$. The number of inserted elements can be estimated by $\hat{D}^d$, where $\hat{D}$ is the average degree of the data graph\cite{chazelle2004bloomier}.  A Bloom filter based path filtering allows us to control the false positives at a low rate with a compact storage and an efficient access time. Moreover, it has no false negatives.

To insert a path into the Bloom filter, we concatenate the labels in the path to form a string that is inserted into the Bloom filter.  To encode the direction of an edge, a sign symbol is added to each label to distinguish between incoming and outgoing edges. In addition, the count of each path is described by preceding the label sequence and separated from the rest of string by ``P''. For example, the string ``2P+1-2'' describes two paths, one with an outgoing edge labeled $1$ and one incoming edge labeled $2$. Two non-matching paths can have $1$ to $d$ unmatched labels.
To avoid filtering out false negatives, we take the lower-bound edit cost $1$ for each non-matching path.

Our experiments show that the two filtering schemes work nicely, complementing each other. Our path-based filtering can identify if multiple labels are in the same path and if the matching edges with the same labels have the same direction, an area the neighborhood-based pruning fails. On the other hand, our neighborhood-based pruning can identify the level of mismatched labels, which cannot be done by our path-based filtering.

\subsection{The WCED Algorithm}

The main problem with EXED is the large number of intermediate results due to backtracking when searching for a mappings, especially for large edit distance thresholds and large node degrees of the data graph. Most of those intermediate results need to be kept until a very late stage of the searching.

To reduce the number of intermediate results, we develop a new algorithm referred to as wildcard queries with edit distance constraint (WCED). The approach taken in this algorithm is to map a subgraph edit distance problem instance into a set of subgraph isomorphism problem instances without missing any relevant answers. This is done by introducing wildcard labels.
A {\it wildcard label} is a label that can substitute for any other label in graph matching.
The main idea is to perform multiple subgraph isomorphism searches based on the original query and merge the retrieved answers to obtain the final results. This approach has two phases: (1) query pre-processing, and (2) subgraph search and answer merging.

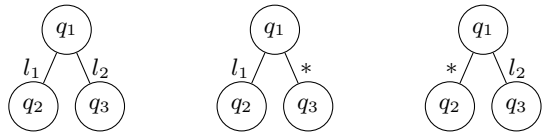
\begin{figure}[tb]
\centering
\input{figures/forest-wcq.tex}
\caption{Query graph and its wildcard queries}
\label{fig:wcq}
\end{figure}

In the query preprocessing phase, we choose $t$ edges from $|E_Q|$ edges in the query graph, where $t$ is the edit distance threshold, and apply an edit operation to each selected edge. Our edit operations are edge label substitution and edge deletion (edge insertion is not considered as discussed in Sec~\ref{sec:def}). For edge label substitution, we set the edge label to wildcard, and for edge deletion, we simply delete the edge. A caveat with edge deletion is that deleting an edge can result in a disconnected query graph, hence deletion can be applied to a subset of the edges while keeping the query connected (see Def.~\ref{def:eteqQ}), whereas substitution can be applied to all edges.
Applying these edit operations gives $O({|E_Q| \choose 2t})$ queries with possibly some wildcards on edge labels, assuming $t \leq |E_Q|$. For example, Figure~\ref{fig:wcq} shows a two-edge query and its wildcard queries with edit distance threshold $1$.

In the next phase, we run subgraph isomorphism searches on those generated queries, where we directly adopt EXED with edit threshold set to $0$. This returns the subgraphs where the wildcard matches any label. For example, searching for the left wildcard query in Figure~\ref{fig:wcq} will give us all subgraphs which have an edge labelled $l_1$ and and an edge with any label, both under the same parent node. Finally, duplicates due to possible overlaps between wildcard queries are removed.

The WCED algorithm reduces the number of intermediate results by converting the subgraph edit distance into subgraph isomorphism. This is for the cost of running EXED ${|E_Q| \choose 2t}$ times with edit distance threshold $0$.



%% file: figures/forest-falsep.tex
\begin{forest}
for tree={edge = {->}, circle,draw, l sep=10pt}
[$q_1$, 
    [$q_3$, edge label={node[midway,left] {$l_1$}}
      [$q_4$,edge label={node[midway,left] {$l_2$}} ] 
      [$q_5$, edge label={node[midway,left] {$l_3$}}] 
      [$q_6$, edge label={node[midway,right] {$l_4$}}]
    ]
    [$q_2$, edge label={node[midway,right] {$l_5$}}] 
]
\end{forest}
\hspace*{1cm}
\begin{forest}
for tree={edge = {->}, circle,draw, l sep=10pt}
[$n_1$, 
    [$n_2$, edge label={node[midway,left] {$l_1$}}]
    [$n_3$, edge label={node[midway,right] {$l_5$}}      
    [$n_4$,edge label={node[midway,left] {$l_2$}} ] 
      [$n_5$, edge label={node[midway,left] {$l_3$}}] 
      [$n_6$, edge label={node[midway,right] {$l_4$}}]] 
]
\end{forest}

%% file: figures/forest-wcq.tex
\begin{forest}
for tree={edge = {->}, circle,draw, l sep=10pt}
[$q_1$, 
    [$q_2$, edge label={node[midway,left] {$l_1$}}]
    [$q_3$, edge label={node[midway,right] {$l_2$}}] 
]
\end{forest}
\hspace*{1cm}
\begin{forest}
for tree={edge = {->}, circle,draw, l sep=10pt}
[$q_1$, 
    [$q_2$, edge label={node[midway,left] {$l_1$}}]
    [$q_3$, edge label={node[midway,right] {$*$}}] 
]
\end{forest}
\hspace*{1cm}
\begin{forest}
for tree={edge = {->}, circle,draw, l sep=10pt}
[$q_1$, 
    [$q_2$, edge label={node[midway,left] {$*$}}]
    [$q_3$, edge label={node[midway,right] {$l_2$}}] 
]
\end{forest}

%% file: sections/05_cost.tex
\section{Algorithm Cost Analysis}
\label{sec:costanalysis}
To better understand the behaviour of our algorithms under different query sizes and edit distance thresholds and to gain some insight on 
which algorithm is expected to perform better for a given query and data graph (without actually running the algorithms), one needs an accurate cost estimation. 

At a high level, WCED decomposes each query into a few wild card queries with the edit distance threshold set to zero for each, whereas EXED runs the same query at most $t+1$ times, each time with a different starting node. More specifically,
EXED has three steps: starting node selection, neighborhood-based pruning and subgraph verification.  The time cost of starting node selection and  neighborhood-based pruning are linear in the number of query nodes and the number of data graph nodes respectively, while the time cost of subgraph verification grows exponentially with the edit distance threshold and the number of query edges. WCED consists of three phases: query pre-processing, subgraph isomorphism search and answer merging. Subgraph isomorphism search uses EXED  with the edit distance threshold $0$, the cost of which also grows exponentially with the number of query edges. The time cost of query pre-processing depends on the number of query edges and the edit distance threshold. The time cost of  answer merging is linear in the number of answers. Both of them are relatively low and negligible compared to the cost of subgraph isomorphism search. Therefore, we focus on the verification cost for both algorithms, which depends on the number of data nodes (candidates) matching the query starting node and the cost of verifying each candidate.

%
\subsection{A Cost Model}
\label{sec:exactCostModel}
Both algorithms EXED and WCED start with a set of candidate nodes in data graph $G$ that are likely to match a query node $n_s$; those candidates may be selected based on a filtering scheme such as the neighborhood-based or the path-based filtering.
Given a candidate node in $G$, we must check if there is a subgraph in $G$ that simulates the query graph in which the candidate node matches $n_s$.
The cost of this process depends on two factors: the number of candidates matching the query node and the cost of verifying each candidate.

\para{A cost model for WCED}
Given a query and an edit distance threshold that is larger than zero,
the WCED algorithm generates a set of wildcard queries based on the edit distance threshold, hence it has to perform multiple subgraph isomorphism searches on those wildcard queries.
The cost is the sum of the costs of those searches.
Note that a wildcard query is like any query except that some edges are labeled with wildcards and those wildcards can match any label.

\noindent {\it Estimating the number of candidates}
Given a seed $n_q$, we estimate the probability that a data node is a candidate mapping for $n_q$.

\begin {lemma}
\label{lemma:pl1-k}
Given a query node and its adjacent edge labels $l_1, \ldots, l_k$, and assuming independence between the labels, the probability that a data node with $D$ adjacent labels has all query labels is
\begin{align}
\label{eq:pl1-k}
\small
&P_D(l_1, l_2, ..., l_k) =  \\ \nonumber
&\sum_{j=1}^{k-1} \sum_{i=j+1}^k (-1)^{i-1}P_D(\neg l_j,\ldots,\neg l_{i-1}, l_{i+1}, \ldots, l_k) +\\ \nonumber
&\sum_{j=1}^{k-1} (-1)^{k-j+1}(1-\sum_{i=j}^k Sel(l_i))^D + (1-(1-Sel(l_k))^D.
\end{align}
\end{lemma}
\begin{proof}
\ifdefined\Archive
  See Appendix~\ref{sec:proofs}.
\else
  \begin{proof}See the extended
 version~\cite{eteq2017}.\end{proof}
\fi
\end{proof}

Lemma~\ref{lemma:pl1-k} directly gives the selectivity of a query node based on its 1-neighborhood.
Let $L_i(n_q)$ denote the set of labels at the $i^{th}$ neighborhood of a query node $n_q$.
The probability that the neighborhood of a data node matches that of a query node at levels $1, \ldots, d$
can be written as
\begin{equation}
\label{eq:plm1-d}
\small
P(n_q) =\prod_{m = 1}^dP_{D_m}(L_m(n_q)),
\end{equation}
where $P_{D_m}(L_m(n_q))$ is as defined in Lemma~\ref{lemma:pl1-k}, and $D_m$ is the number of edges at the $m^{th}$ neighborhood of a data node.
We generally do not know $D_m$ when estimating our probabilities in Equations~\ref{eq:pl1-k}.
Assuming that each data node has the same degree $\hat{D}$, $D_m=\hat{D}^m$.
Then, the number of candidates matching query node $n_q$ is
$|C(n_q)| = |V| * P(n_q)$.

\noindent {\it Estimating the cost of verifying each candidate}
For each candidate of the starting node, the algorithm starts from a graph with only one node (i.e. the candidate node) and iteratively adds new edges to it until either the resulting graph simulates the query, or no such simulation is found.
The cost of adding each new edge depends on the expected number of matching edges of a query edge and the number of subgraphs to which the edges are added.
For a query label $l_i$, we expect $\hat{D}*Sel(l_i)$ edges of a node in the data graph to match $l_i$.
For a fixed candidate node in the data graph, the expected number of subgraphs (partial matchings) that can be constructed starting from the candidate and simulating the query subgraph rooted at the seed with labels $l_1, \ldots, l_k$ is
$\prod_{i=1}^{k}\hat{D}*Sel(l_i)$ 
and the total expected cost of verifying a candidate $n$ is
\begin{equation}
\label{eq:cost-verifying-n}
\small
\sum_{i=1}^{|E_q|}\prod_{j = 1}^i\hat{D}*Sel(l_j).
\end{equation}
Note that this is based on the assumption that a search starting from a candidate node will not stop early if the simulation exceeds the edit distance threshold.
The total cost of verifying $|C(n_q)|$ candidates is
\begin{equation}
\label{eq:cost-verifying-q}
\small
Cost(q) = |C(n_q)| * \sum_{i=1}^{|E_q|}\prod_{j = 1}^i\hat{D}*Sel(l_j).
\end{equation}

Since we have replaced a query graph with ${|E_q| \choose 2t}$ graphs each with $t$ wildcards, the total cost is the sum of the costs of verifying those wildcard queries.


%
\ifdefined\Archive
\noindent{\bf EXED Cost Model}
To estimate the EXED cost, we also need to estimate the number of candidates in the data graph matching a query seed node, and the cost of verifying each candidate.

\noindent {\it Estimating the number of candidates}
Since a data node is allowed to have up to $t$ edit operations in its neighborhood, directly estimating the probability that a data node is a qualified candidate is difficult. Therefore, with a fixed starting node $n_q$, we estimate the number of candidates for a set of wildcard queries where the labels are all fixed.
By summing up the number of candidates for these wildcard queries and removing the repetitive candidates due to overlaps between queries,  the number of candidates for $n_q$ in EXED can be written as
\begin{align}
\label{eq:Cnq}
\small
|C(n_q)| &= \sum_{i=1}^{|E_q| \choose 2t}|V_g| * P(n_{w_i(q,t)}) \nonumber\\
&- ({|E_q| \choose 2t} - 1) * |V_g| * P(n_q),
\end{align}
where $w_i(q,t)$ is a wildcard query constructed from $q$ by replacing $t$ edge labels with wildcards
and $P(n_q)$ is as in Equation~\ref{eq:plm1-d}. The last term gives the number of double-count candidates for ${|E_q| \choose 2t}$ wildcard queries.

\noindent {\it Estimating the cost of verifying each candidate}
To estimate the cost of verifying each candidate, we need to estimate the number of partial matches.
There are two kinds of partial matchings in EXED: (1) matchings that have reached the edit distance threshold, and (2) matchings that have not reached the threshold. For (1), edges with any label can be added to the matching in the next step of the simulation, whereas for (2), only edges with matching labels can be added. Let $m$ be the number of edges in a partial matching, and $k$ be the number of edges in a matching where the matching edges have different labels. If $l_1, \ldots, l_k$ denote the query labels in the matching where the labels don't match, and $l_{k+1}, \ldots, l_m$ be the
labels where both data and query edges in the matching have the same labels,
then the number of partial matchings can be written as:
$\hat{D}^m \prod\limits_{i=1}^{m-k}Sel(l_i)\prod\limits_{j=1}^{k}(1-Sel(l_j))$.
Given query labels $l_1, \ldots, l_m$, we generally do not know in advance which labels will mismatch and need to check all choices of ${m \choose t}$ sets of labels. The number of partial matchings that need to be verified is
\begin{equation}
\label{eq:st-cases}
\small
    S_t(q,m)=
\begin{cases}
    0         			  & \text{if } t > m\\
    \hat{D}^m\prod\limits_{i=1}^{m}Sel(l_i) & \text{if }t=0\\
    \sum\limits_{k=1}^{m \choose t}\hat{D}^m \prod\limits_{i=1}^{m-t}Sel(l_{k,i})\prod\limits_{j=1}^t(1-Sel(l_{k,j}))& \text{if } t < m.
\end{cases}
\end{equation}

For any partial matching that have not reached the threshold $t$, any edge can be added into the matching in the next step of the simulation. In this case, the next step of simulation costs:
$\sum\limits_{j=0}^{t-1}S_j(q,m) * \hat{D}$.

For any partial matching that have reached the threshold, only edges with a matching label can be added. In this case, the next step of a simulation costs
$S_t(q,m) * \hat{D} * Sel(l_{m+1})$.

The cost of verifying each candidate in  EXED is
\begin{equation}
\label{eq:ed-verify}
\small
Cost(q) = \sum\limits_{i=0}^{|E_q|-1}(S_t(q,i) * \hat{D} *  Sel(l_{i+1}) + \sum\limits_{j=0}^{t-1}S_j(q,i) * \hat{D})
\end{equation}
\noindent and the total cost of  EXED is the product of the number of candidates (as given in Equation~\ref{eq:Cnq}) and the cost of verifying a candidate (as given above):
$Cost_{ex} = |C(n_q)| * Cost(q)$.
\else
\para{EXED Cost Model.}
A similar cost model can be developed for EXED by estimating the number of candidates in the data graph matching a query seed node, and the cost of verifying each candidate. 
 For more details see the extended version~\cite{eteq2017}.
\fi

\para{Cost Model Comparison.}
We compare the costs of verifying the candidates for EXED and WCED and identify the conditions under which one outperforms the other.
Our comparison assumes that the threshold $t$ is less than the number of query edges; otherwise, the problem is subgraph isomorphism with no label constraints, which is not addressed in this paper.

For an edit distance threshold larger than zero, the cost of verifying a candidate in EXED is higher than that in WCED, because edit operations can happen on any label in EXED while the labels are all fixed in WCED. Hence if the number of candidates for WCED and EXED are roughly the same, WCED will outperform EXED. In other words, WCED outperforms EXED if the number of candidates for the original query is small. 
This is a more plausible scenario for our queries; otherwise edit operations are less likely to be considered. The next lemma shows what happens when this condition does not hold.


\begin {lemma}
\label{lemma:comp}
Given a data graph with expected node degree $\hat{D}$, a query graph $q$ with at least $2$ edges and the edit distance threshold set to $1$,
the cost of verifying a candidate in EXED is less than the sum of the cost of verifying a candidate for every wildcard queries in WCED when
$Sel(l_1) > 1/ \sqrt[|E_q|]{\hat{D}}$,
where $l_1$ is a query label that has the highest selectivity (i.e. the smallest value of $Sel(l_i)$).
\end{lemma}
\begin{proof} 
\ifdefined\Archive
  See Appendix~\ref{sec:proofs}.
\else
  See the extended
 version~\cite{eteq2017}.
\fi
\end{proof}
When the number of candidates for the original query is large (roughly equal to the number of candidates for a wildcard query), the cost of EXED and WCED can both be approximated based on  the number of candidates for the original query. In this case, EXED can outperform WCED given the condition of the lemma.

%

\subsection{An Upper Bound Cost Model}
\label{sec:ub}
The cost model discussed in the previous section is based on assumptions that labels are both evenly distributed and pairwise independent. These assumptions may not hold in real-world data graphs. This is a problem especially for large queries since the error can accumulate and become significant as the number of query edges increases. 
In this section, we present a cost model that gives an upper bound of the actual cost but is more accurate for larger query graphs.

\para{Estimating the number of candidates.}
To estimate an upper bound on the number of candidates, two weaker assumptions of label independence are considered: (1) the labels of the adjacent edges of a data node are independent whereas labels, which are in a path starting from a node, are correlated; (2) the labels of the adjacent edges of a data node are correlated whereas labels, which are in a path starting from the node, are independent. For two or more correlated labels, the selectivity of the label with the least selectivity provides an upper bound of the selectivity of the set.

Under the first assumption, the selectivity of the label with the minimum selectivity in each path is used  to estimate the selectivity upper bound of the path. This reduces each path in the query to an edge (with the minimum selectivity), and as a result the query becomes a node with a set of adjacent edges (i.e. a tree with only one level). Assuming independence between the labels of these edges, Lemma~\ref{lemma:pl1-k} will give an upper bound of  the probability that a data node is a candidate for a query node. Note that $D$ in the Lemma is set to the number of paths in the $d$-neighborhood.

Under the second assumption, all edges under a node are collapsed into a single edge, which is labeled with a label from the set that has the least selectivity. Since the edge labels of the resulting query are all independent, Equation~\ref{eq:plm1-d} can be used to estimate the upper bound.


\para{Estimating the cost of verifying each candidate.}
To estimate an upper bound on the cost of verifying each candidate, the maximum frequency of each label under a node is used to upper bound the number of matching label in each step of the simulation.
Let $N(l_i)$ denote the maximum frequency of label $l_i$ in the adjacent edges of a node.
In our cost model, the number of matching labels for a  label $l_i$ is $\hat{D}*Sel(l_i)$ assuming that every label is uniformly distributed on the adjacent edges of a node. Replacing $\hat{D}*Sel(l_i)$ in the cost estimates (i.e., Equation~\ref{eq:cost-verifying-q} for WCED 
\ifdefined\Archive
and Equation~\ref{eq:ed-verify} for EXED
\fi
)
by $N(l_i)$ 
will give us an upper bound of the cost of verifying each candidate in WCED and EXED.

%% file: sections/05p_extensions.tex
\section{Refining Match Constraints}
\label{sec:extensions}
Error tolerant exemplar queries with its leverage of edit operations are quite powerful for exploratory searches and can retrieve many matches that otherwise cannot be found under an edge isomorphic match. That said, it is not hard to show that the match condition in ETEQ can further be refined to better support exploratory searches.

\para{Matching node labels.}
Consider the example of finding the founders of technology companies (presented in Section~\ref{sec:intro}), and suppose we want to retrieve companies listed under the Information Technology (IT) industry. In other words, instead of completely ignoring node labels, we want to match the labels on a selected set of nodes. One way to support such matches is in post-processing, after retrieving ETEQ matches and before returning the result to the user. A more efficient solution is to push the constraint on node labels into the ETEQ evaluation engine and make our index structures aware of the node labels. To incorporate node labels in our neighbourhood index, one can map any graph with labels on both edges and nodes to a graph that has labels on edges but not nodes. Suppose node and edge labels are disjoint; if not, one can add a prefix (e.g. \$) to node labels to make them disjoint. For each node $n$ with label \textit{l}, create a dummy node, say $n^\prime$, and an edge labelled \textit{l} between $n$ and $n^\prime$. The direction of the edge is not important as long as they are consistent, say all such edges are from graph nodes to the dummy nodes.

A neighborhood index on this extended graph, constructed as discussed before, will allow searches on both edge and node labels. For searches on edge labels only, the new edges in the index will not introduce any false positives since the node and edge labels are disjoint. For queries with constraints on both node and edge labels, the constraints on node labels can be mapped to constraints on edge labels as discussed above and they can be pushed to the index to further prune the search space. Our path index may be revised to incorporate node labels as well; this revision is not discussed here for brevity.

\para{Semantic matching of edge lables.}
As another refinement, semantic relationships between edge labels may be incorporated in the edit distance measure. It is reasonable to assume that the cost of a semantically related match (e.g., matching synonyms) is not larger than that of an edit operation (e.g., edge substitution), hence semantic matching can be implemented as a post-processing step by ordering the matches that are within the same edit distance, based on a measure of semantic similarity between edges that are aligned in the match but have different labels.
In line with other works~\cite{zheng2016semantic,elbassuoni2011query,huang2012approximating} we can exploit a similarity measure between edge labels, so that given 2 edge labels we can establish whether they have a similar meaning (e.g., to establish that \emph{author of} is more similar to \emph{creator of} than to \emph{owner of}).
Starting from answers at distance 1 and then moving to larger distances, we can re-weight the edit-distance of answers based on the label that is swapped.

%% file: sections/06_experiments.tex
\section{Experimental Evaluation}
\label{sec:experiments}
This section presents the experimental evaluation of our algorithms and cost models. All experiments were performed on a 2.4GHz 4 Core CPU with 60G RAM running Linux. The algorithms are implemented in Java 1.8. Unless stated otherwise, the path length $d$ in our filtering scheme is $3$.

\para{Summary of findings.}
In the following, we demonstrate the advantages in terms of scalability and in term of answer quality of ETEQ compared to other graph search methods (Sec~\ref{sec:comp}).
In particular, we compare ETEQ to both the original (non-approximate) graph exemplar query solution and to  another state of the art algorithm for approximate graph search.

Moreover, we demonstrate how the proposed cost model provides accurate cost estimations and thus allows us to select the best algorithm (Section~\ref{sec:cost}); in particular the \emph{upper bound model} provides the best performances for queries with more than 2 edges.
Moreover, we demonstrate how the proposed pruning strategies are able to prune between $78\%$ and $99\%$ of data nodes (Sections~\ref{sec:filtering}), depending on the complexity of the query, and that by applying both strategies together can provide effective pruning and limit the number of false positives (Section~\ref{sec:comb}).

\para{Dataset.}
We downloaded a full dump of Freebase\footnote{\url{https://developers.google.com/freebase}} as of May 2015 and removed the triples that were used as internal specification for the community (e.g., user and group data and discussion topics), obtaining a fully connected graph of $84$ million nodes and $335$ million edges.
Since the entire graph of Freebase requires at least $90$G of memory when fully loaded, given our computational resources, we extracted subgraphs from Freebase with different parameters, using a breadth first traversal of the graph from a randomly selected starting node and randomly choosing new edges to be included in the data graph.
Unless explicitly stated otherwise, the data graphs are randomly generated from Freebase with the number of nodes set to $10$K and the average node degree set to $15$.

\para{Queries.}
Three types of queries are used in our experiments: (1) a set of real queries from the AOL query log, manually mapped to the data graph, (2) a set of real queries from QALD-4~\footnote{\url{http://qald.sebastianwalter.org/index.php?x=challenge&q=4}}, a benchmark for evaluating question answering over linked data, and (3) randomly selected subgraphs of the data graph.
These queries vary in their number of edges and the selectivity of their labels.
The AOL and QALD-4 queries were used to evaluate the quality of the answers retrieved by ETEQ.
While, unless explicitly stated otherwise, our experiments use $100$ randomly selected queries, each a subgraph of the data graph, to test the performance of ETEQ and its components under various conditions.

\subsection{Compared to Competitors}
\label{sec:comp}
To evaluate our algorithms against competitors, we selected two state-of-the-art algorithms from the literature: (1) SAPPER~\cite{zhang2010sapper}, which is proposed for indexing and approximate matching in large graphs, and (2) the original exact Exemplar Query method~\cite{mottin2014exemplar}, which is similar to our work but limited to edit distance zero.

\begin{figure*}[tb]
\centering
\begin{minipage}{.3\textwidth}
\hspace*{-0.5cm}
\includegraphics[width=2.35in]{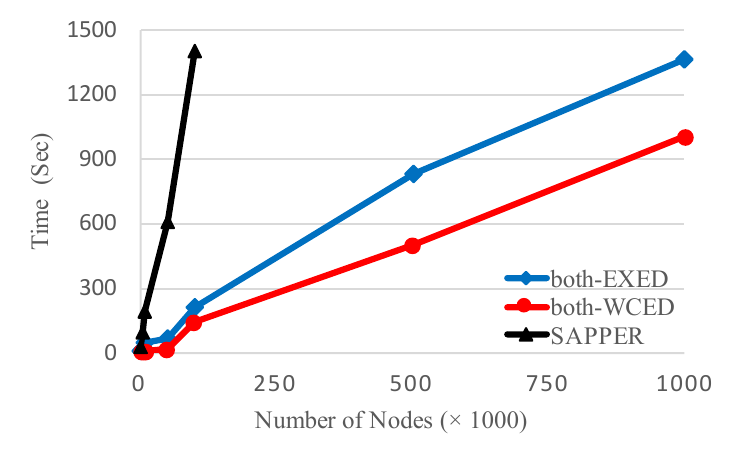}
\caption{Running time for different algorithms}
\label{fig:comp}
\end{minipage}
\hspace*{0.3cm}
\begin{minipage}{.3\textwidth}
\hspace*{-0.5cm}
\includegraphics[width=2.35in]{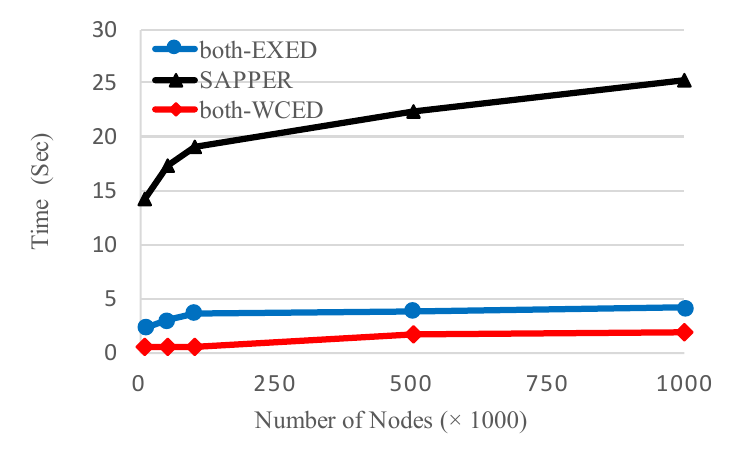}
\caption{Running time for fetching the first 100 answers}
\label{fig:comp_100}
\end{minipage}
\hspace*{0.3cm}
\begin{minipage}{.3\textwidth}
\hspace*{-0.5cm}
\includegraphics[width=2.35in]{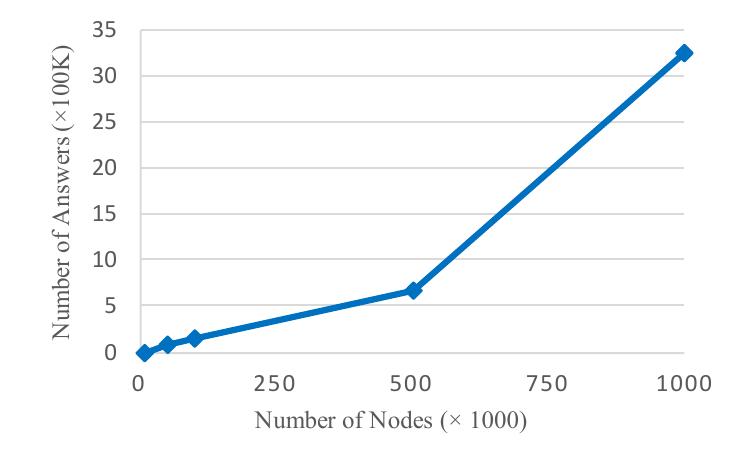}
\caption{Answer size varying the number of nodes in data graph}
\label{fig:answer_num}
\end{minipage}
\end{figure*}

\para{Compared against SAPPER.}
In this set of experiments, we compare the scalability of our algorithms against SAPPER. In the first experiment,
we varied the number of nodes in the data graph from $10$K to $1$M while the edit distance threshold was set to $1$. This is consistent with the settings by the authors of SAPPER except that their largest data graph had only $10$K nodes. Since SAPPER only supports edge deletion (missing edges), we modified SAPPER to support edge label substitutions.
Figure~\ref{fig:comp_100} shows the running time for retrieving the first 100 answers.
The results show that SAPPER is slower by an order of magnitude, and it becomes slower as the data graph size increases.
In contrast, our algorithms are not much sensitive to the size of the data graph and scale gracefully to large data graphs, exhibiting an almost constant behavior.


In our second experiment, we used the same setting as the first one, except that we did \emph{not} limit the number of answers; instead we set a 1500 seconds time limit on each algorithm.
The running times for SAPPER and our algorithms are reported in Figure~\ref{fig:comp}.
The graph shows that SAPPER is not a viable solution to our problem for data graphs of realistic sizes: the running time of SAPPER grows much faster than our algorithms, and quickly hits the 1500 seconds time limit in a data graph with $100$K nodes. WCED with both filtering schemes exhibits the best performance. We also observe that the time cost of our algorithms grows linearly with the number of nodes in the data graph. This meets our expectation, because the number of relevant answers increases rapidly (See Figure~\ref{fig:answer_num}) as the number of data nodes increases.

%

\begin{figure*}[tb]
\centering
\begin{minipage}{0.33\linewidth}
\includegraphics[width=\textwidth]{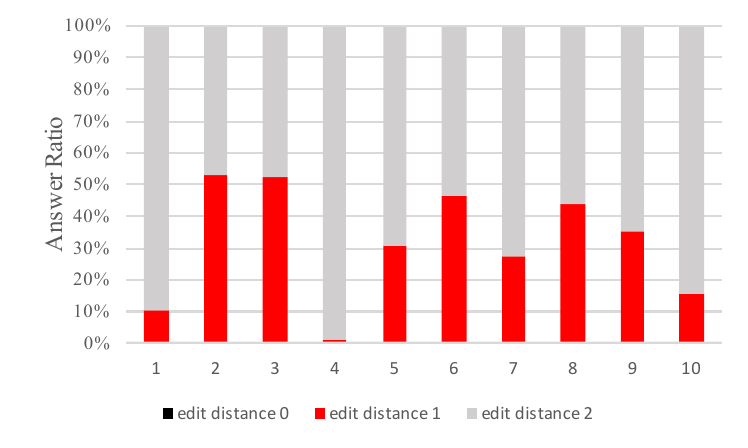}
\caption{Answer set composition}
\label{fig:exqcomp}
\end{minipage}
\begin{minipage}{0.66\linewidth}
\subfigure{
\centering\includegraphics[width=0.5\textwidth]{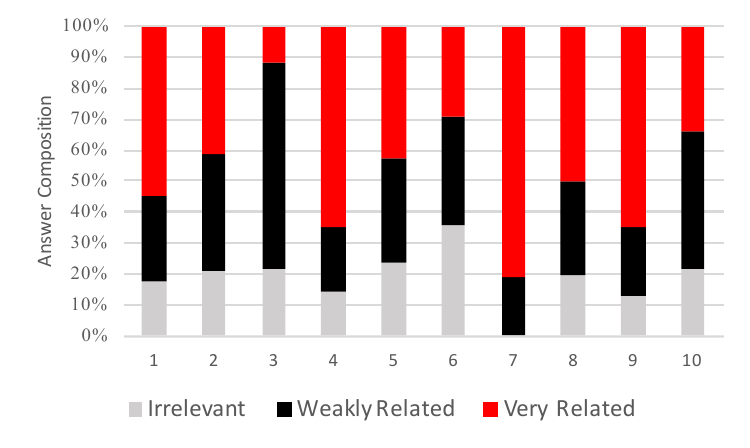}
}
\subfigure{
\centering\includegraphics[width=0.5\textwidth]{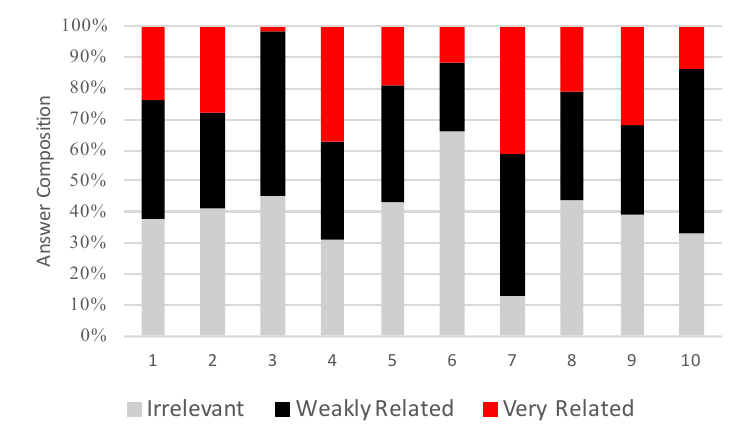}
}
\caption{Relevant answer set composition for edit  distance $1$ (left) and $2$ (right)}
\label{fig:edits}
\end{minipage}
\end{figure*}

\para{Compared against Exemplar queries.}
To compare the effectiveness of our queries to that of exemplar queries, we ran experiments using queries from both AOL query log and the QALD4 benchmark. From the AOL query log,
we chose $10$ queries (see 
\ifdefined\Archive
  Appendix~\ref{sec:query_set}
\else
  the extended version of this paper~\cite{eteq2017}.
\fi
) and manually mapped them to Freebase.
For each query, we introduce errors by randomly selecting an edge and replacing its label with a label randomly selected from the data graph.
The number of query edges ranged between $6$ and $8$. 
To control the size of the answer set (and to avoid a blow-up), we varied the edit distance threshold from $0$ to $2$.
When the edit distance threshold is $0$, our queries are identical to exemplar queries.
As expected and shown in Figure~\ref{fig:exqcomp}, exemplar queries fail to return any answer for queries with errors, whereas ETEQ retrieves the answers despite the error, and the larger the edit distance thresholds are, the more answers are returned.


To evaluate the quality of ETEQ answers, we conducted the following user study.
We asked 10 users (students at the University of Alberta) to evaluate our system.
For each query in the test set, we provided an explanation of the topic, the query intention, and our answer set with different edit distances.
We asked each user to rate each result as irrelevant, weakly related, or very related with respect to the topic and the expressed query intent.
Due to the large size of the answer sets, for each answer set and each edit distance, we randomly chose up to $10$ answers for evaluation.
We observe in Figure~\ref{fig:edits} that the relevant set has many answers with edit distances $1$ and $2$. 
These answers cannot be returned by exemplar queries.

We also evaluated the effectiveness of our ETEQ queries over the QALD4 benchmark, a collection of natural language questions over linked data.
To adapt this benchmark to our framework, we focused on list questions; we selected one answer for each question (the first one given in the benchmark) and collected all predicates from DBpedia that had that answer (as a subject, or an object) and a term from the question.
For example, for the question ``which books by Kerouac were published by Viking Press?'', the predicates  ``$\ast$Kerouac, notableWork, X'', ``X, author, $\ast$Kerouac'' and ``X, publisher, Viking\_Press'' were collected, where X indicates the initial given answer and $\ast$ indicates a wild card.
We also collected pairs of predicates that could be joined, giving a path of length 2, and the path had both the answer and a question word.
For example, for the question ``Give me all actors starring in movies directed by William Shatner,'' the predicates ``a, starring, X'' and ``a, director, William\_Shatner'' were collected.
Our goal was to find more answers matching X from a given example.
Moreover, given the nature of the task, we also extended our algorithm to verify matches on node labels that were present, such as ``$\ast$Kerouac'' and ``Viking\_Press''.
Since the answers to questions were given, we could track at each edit distance the answers that were returned.

As shown in Table~\ref{tbl:qald4} for 20 queries from this benchmark~\footnote{The selected queries were the first set of list queries with $\ge 3$ answers, and excluded more trivial questions that only matched 1 predicate. Both the queries and the predicates can be found in 
\ifdefined\Archive
  Appendix~\ref{sec:query_set}.
\else
  the extended version of this paper~\cite{eteq2017}.
\fi
}, only 37\% of the answers are at edit distance zero and can be returned using exemplar queries, whereas the rest of the answers are at larger edit distances and can only be returned using our ETEQ queries.

We are not comparing the efficiency of our algorithms against Exemplar Queries~\cite{mottin2016exemplar}, because (1) their framework does not support edit distance thresholds larger than zero, and (2) our EXED algorithm becomes identical to their exemplar queries when the edit distance threshold is zero, and we have extensively evaluated EXED with different edit distance thresholds.
Also since it is shown that exact exemplar queries outperform NeMa~\cite{khan2013nema}, we are not comparing to NeMa.

\begin{table}[tb]
\begin{tabular}{lccccc}
Edit distance & 0 & 1 & 2 & 3 & 4\\ \hline
\% of answers (mean) & 0.37 & 0.23 & 0.25 & 0.06 & 0.04 \\
\% of answers (std) & 0.35 & 0.25 & 0.32 & 0.11 & 0.10\\
\end{tabular}
\caption{The fraction of answers at each edit distance for 20 queries from QALD4 benchmark.}
\label{tbl:qald4}
\end{table}

\subsection{Effectiveness of Our Cost Models}
\label{sec:cost}
\begin{figure}[tb]
\centering
\centering\includegraphics[width=2.7in]{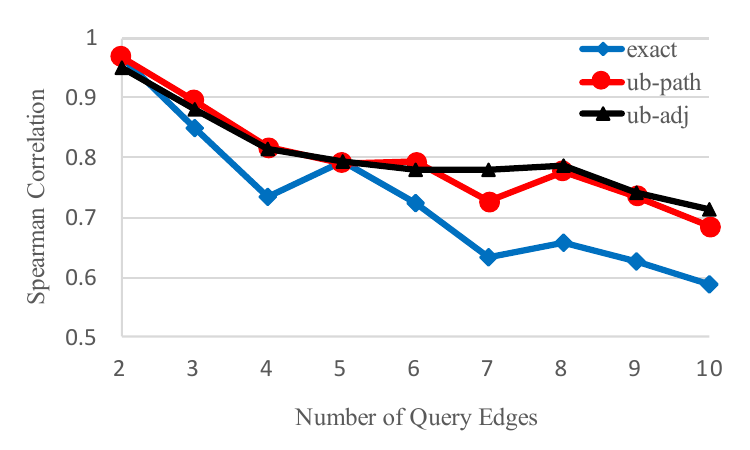}
\caption{Correlation between estimated and actual selectivities for WCED}
\label{fig:sels}
\end{figure}

%
%
\para{Effectiveness of the selectivity estimation}
Figure \ref{fig:sels} shows the correlation between our selectivity estimates and the actual selectivitity for WCED, measured in terms of the Spearman's rank correlation, which shows the monotonic relationship between the variables. In our case, the selectivity is  used in choosing a query starting node and for cost comparisons, hence, a relative ordering of the selectivity values is sufficient. In the figure, ``exact'' denotes our selectivity estimate under the independence assumption (as discussed in Section~\ref{sec:exactCostModel}), whereas ``ub-path'' and ``ub-adj'' denote the upper bounds of the selectivity estimations respectively  assuming that labels in a path and labels of edges adjacent to a node are independent. Both exact and upper bound estimates perform well (correlation over 0.55) for a large range of query sizes, with exact performing better (correlation $0.96$) for queries that have 2 edges whereas the upper bound performing better for larger query sizes. Similar result was observed for EXED (not shown for brevity).

\begin{figure*}[tb]
\centering
\begin{minipage}{.3\textwidth}
\hspace*{-0.5cm}
\includegraphics[width=2.35in]{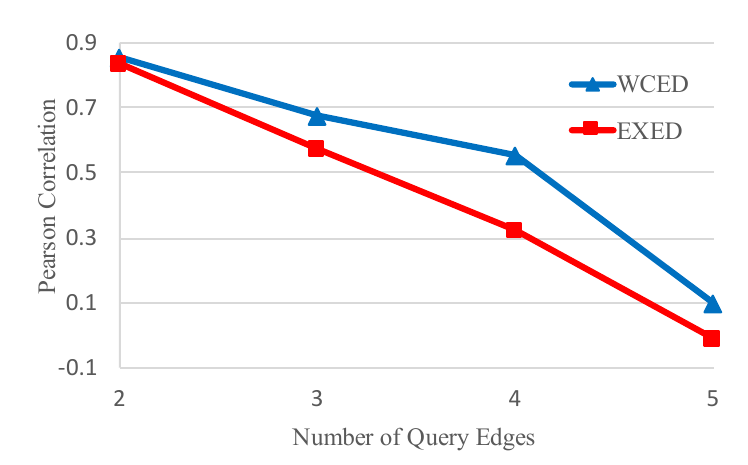}
\caption{Correlation between actual and estimated cost differences varying the number of query edges}
\label{fig:exact-cost}
\end{minipage}
\hspace*{0.25cm}
\begin{minipage}{.3\textwidth}
\hspace*{-0.5cm}
\includegraphics[width=2.35in]{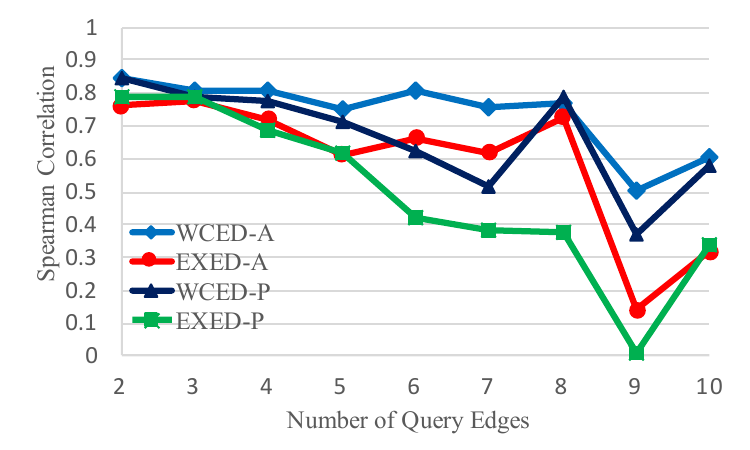}
\caption{Correlation between estimated and actual costs varying the number of query edges}
\label{fig:sel}
\end{minipage}
\hspace*{0.25cm}
\begin{minipage}{.3\textwidth}
\hspace*{-0.5cm}
\includegraphics[width=2.35in]{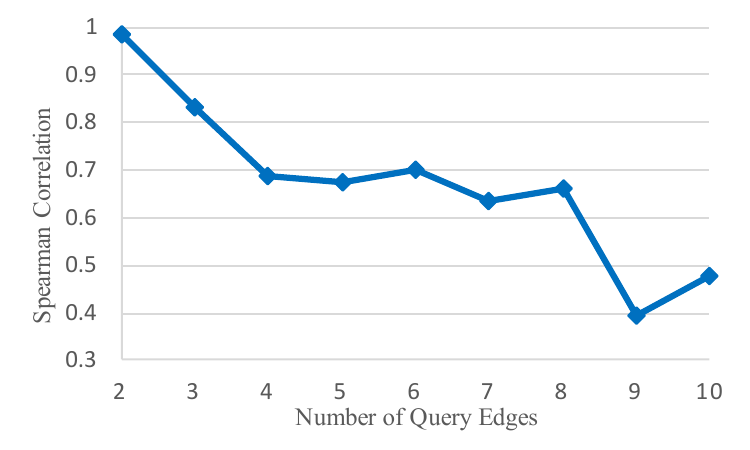}
\caption{Correlation between actual and estimated cost differences varying the number of query edges}
\label{fig:costcomp}
\end{minipage}
\end{figure*}

\para{Cost model evaluation}
In this set of experiments, we examine the linear relationship between our estimated cost  and the actual number of operations using Pearson correlation. The larger the absolute value of the coefficient, the stronger the relationship between the actual cost and the estimated cost. For relatively large values of the correlation coefficient, one can predict with a good accuracy the actual cost from our estimated cost, using a simple linear regression.
Figure~\ref{fig:exact-cost} shows that for small queries (with up to $3$ edges), the correlation coefficient is over $0.7$ and $0.6$ for WCED and EXED respectively. However, the correlation coefficient drops sharply as the number of edges increases. These results are expected because the cost model is based on the assumption that the labels are evenly distributed and that they are independent.
%

\para{Upper-bound cost model evaluation}
In this set of experiments, we evaluate the upper bound cost models of our algorithms presented in Section~\ref{sec:ub}. Let's denote with WCED-A and EXED-A the upper bounds of the cost model of WCED and EXED assuming independence of adjacent labels respectively, and denote with WCED-P and EXED-P the upper bounds of the cost model of WCED and EXED assuming independence of path labels respectively. Figure~\ref{fig:sel} shows that both WCED-A and EXED-A have a better Spearman correlation with the actual cost than both WCED-P and EXED-P. Both WCED-A and EXED-A have over $0.6$ correlation for queries up to $8$ edges, while WCED-P  only has $0.5$ correlation when queries have $7$ edges and the correlation of EXED-P drops below $0.4$ when queries have more than $6$ edges. Based on these results, both WCED-A and EXED-A provide good cost models for comparing the cost of different queries.

To further evaluate the effectiveness of our cost models, we computed the gaps between the actual costs of WCED and EXED, i.e., actW-actE and their estimated costs, i.e., estW-estE. A high correlation between the two gaps indicates that the cost model can show which algorithm has the least cost even though the actual value of the estimate may not be accurate. Figure~\ref{fig:costcomp} shows that for queries with up to $6$ edges, the correlation is strong (over $0.7$).


\begin{figure*}[tb]
\centering
\begin{minipage}{.3\textwidth}
\hspace*{-0.5cm}
  \centering\includegraphics[width=2.35in]{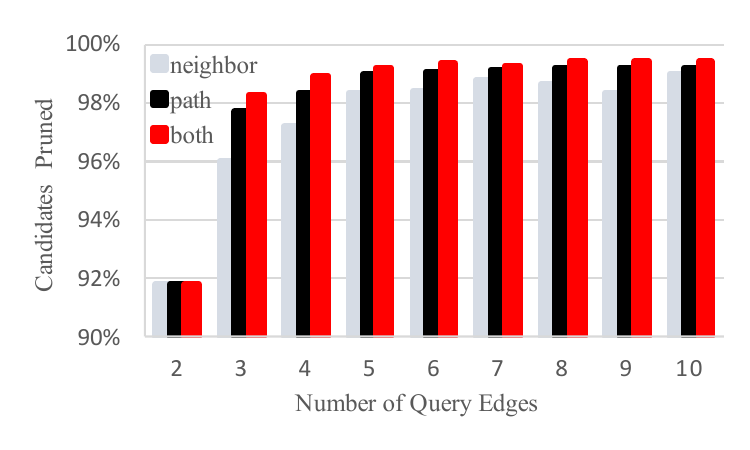}
\end{minipage}
\hspace*{0.3cm}
\begin{minipage}{.3\textwidth}
\hspace*{-0.5cm}
\centering\includegraphics[width=2.35in]{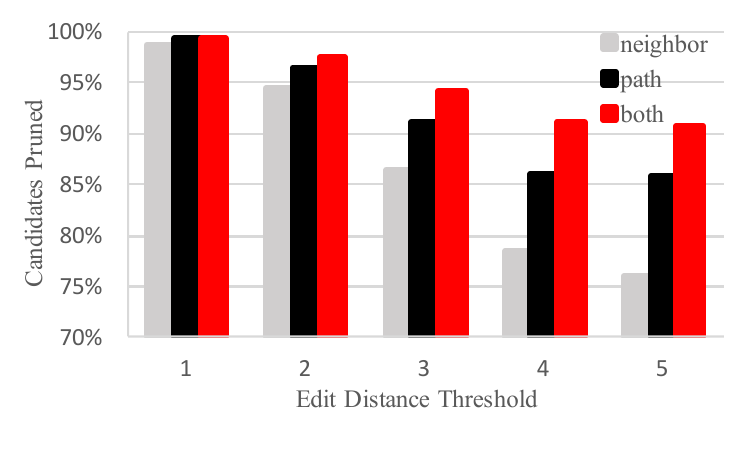}
\end{minipage}
\hspace*{0.3cm}
\begin{minipage}{.3\textwidth}
\hspace*{-0.5cm}
\centering\includegraphics[width=2.35in]{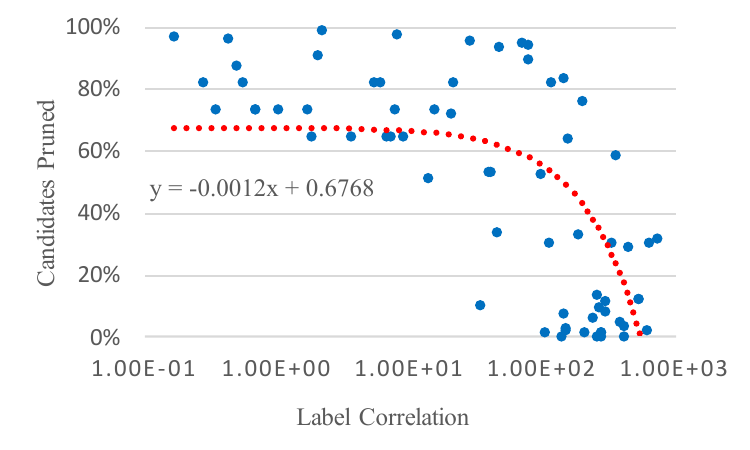}
\end{minipage}
\begin{minipage}{.3\textwidth}
\hspace*{-0.5cm}
  \centering\includegraphics[width=2.35in]{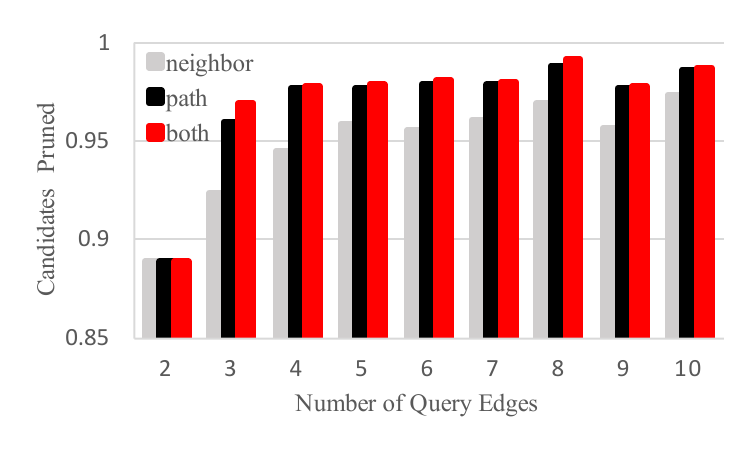}
  \caption{Pruning power of WCED (top) and EXED (bottom) for different number of query edges}
  \label{fig:prune}
\end{minipage}
\hspace*{0.3cm}
\begin{minipage}{.3\textwidth}
\hspace*{-0.5cm}
  \centering\includegraphics[width=2.35in]{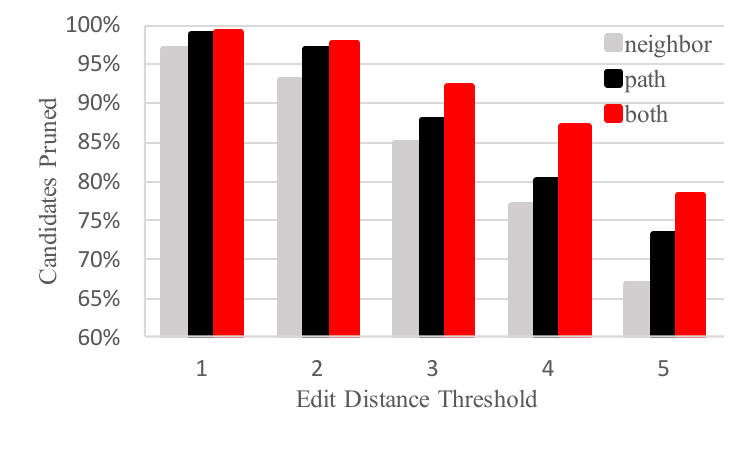}
  \caption{Pruning Power of WCED (top) and EXED (bottom) for different edit distance thresholds}
  \label{fig:threshold}
\end{minipage}
\hspace*{0.3cm}
\begin{minipage}{.3\textwidth}
\hspace*{-0.5cm}
\centering\includegraphics[width=2.35in]{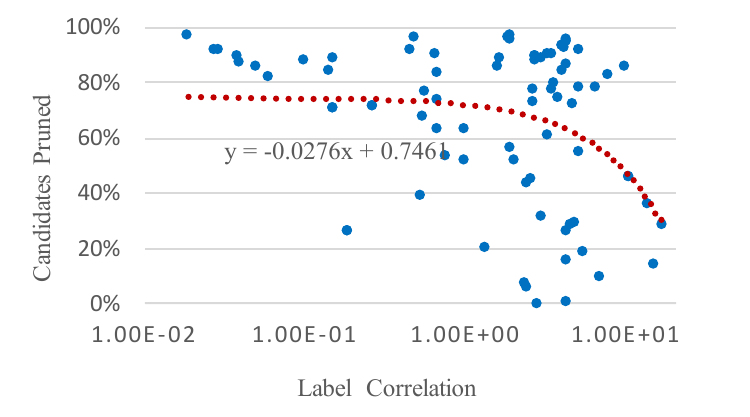}
\caption{Candidates pruned varying path label correlation for WCED (top) and EXED (bottom)}
\label{fig:labelcor}
\end{minipage}
\end{figure*}

\subsection{Effectiveness of Our Filtering Strategies}
\label{sec:filtering}
To evaluate the pruning power of our filtering schemes, the number of nodes in the data graph was set to $10$K. 
Let ``neighbor'' denote the neighborhood-based pruning strategy, ``path'' denote the path-based filtering strategy and ``both'' denote the case where both schemes were used. 
The integration of our pruning strategies into EXED is as discussed in Section~\ref{sec:approach}. WCED calls EXED with the edit distance threshold zero, and our pruning strategies are applied through EXED.


\para{Varying the number of query edges.}
For this experiment, we varied the number of query edges from $2$ to $10$ and set the edit distance threshold to $1$. Figure~\ref{fig:prune} shows the fraction of candidates that are pruned in EXED and WCED as we vary the number of query edges: for WCED and EXED ``path'' can filter out respectively  up to $99.4\%$ and $99.1\%$ of the data nodes on average, while ``neighbor'' can filter out respectively up to $99.0\%$ and  $97.3\%$ of the data nodes on average. The pruning power does not increase by more than $1$\% when both strategies are used. However, considering the large number of data nodes and the high cost of verifying each candidate, even a small improvement in the pruning stages positively affects the performance of the algorithms (Figure~\ref{fig:edgenumber}).


\para{Varying the edit distance threshold.}
In this experiment, the number of query edges was fixed at $8$, and edit distance threshold varied from $1$ to $5$. When the edit distance threshold is equal to or exceeds the number of query edges, the labels become irrelevant and the problem becomes subgraph isomorphism on unlabeled graphs, which we do not address in this paper. Figure ~\ref{fig:threshold} shows that both ``neighbor'' and ``path'' have good pruning power (over $78\%$) under different distance thresholds, and it becomes more effective to apply both filtering schemes as the edit distance threshold increases. This is because ``neighbor'' scheme does not encode edge direction in its indexes and higher edit distance threshold introduces more false positives with wrong edge direction, while adding ``path'' on top of ``neighbor'' can effectively prune out those false positives.


\para{Varying path label correlation.}
For this experiment, the neighborhood-based pruning scheme is considered as a baseline, on top of which we added our path-based filtering scheme and monitored the improvement in pruning power. We fixed  the number of query edges at $8$ and set the edit distance threshold to $1$. As shown in Figure~\ref{fig:labelcor} , the improvement in pruning power by adding ``path'' in both EXED and WCED drops with more correlation. This meets our expectation since the more correlated the labels are, the less false positives the neighborhood-based pruning can produce and the less room for ``path'' filtering improvements.

\subsection{Combining Filtering Schemes}
\label{sec:comb}
In this set of experiments, we evaluate the impact of adding path-based filtering on top of the neighborhood-based pruning. In order to show the impact of using both filtering schemes, we consider EXED with the neighborhood-based pruning scheme as our baseline and compare it against EXED with both filtering schemes, WCED with the neighborhood-based pruning scheme and WCED with both filtering schemes. We denote EXED and WCED with the neighborhood-based pruning scheme as ``neighbor-EXED'' and ``neighbor-WCED'' respectively, WCED and EXED with both filtering schemes as ``both-WCED'' and ``both-EXED'', respectively.

\begin{figure*}[tb]
\centering
\begin{minipage}{.3\textwidth}
\hspace*{-0.5cm}
\includegraphics[width=2.35in]{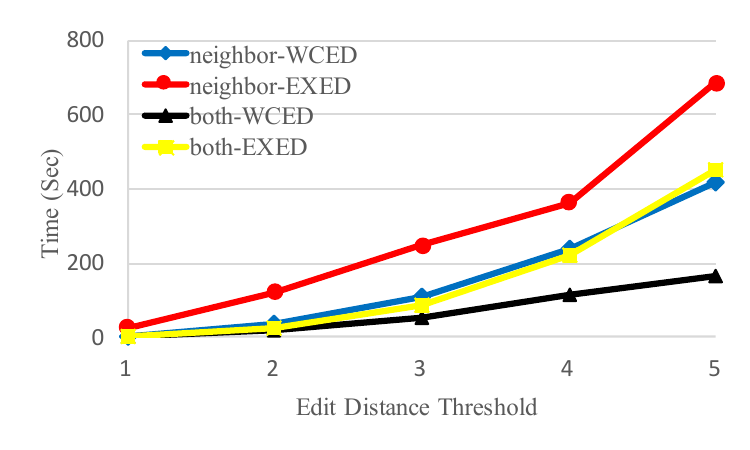}
\caption{Running time varying edit distance threshold}
\label{fig:10ktime}
\end{minipage}
\hspace*{0.25cm}
\begin{minipage}{.3\textwidth}
\hspace*{-0.5cm}
\includegraphics[width=2.35in]{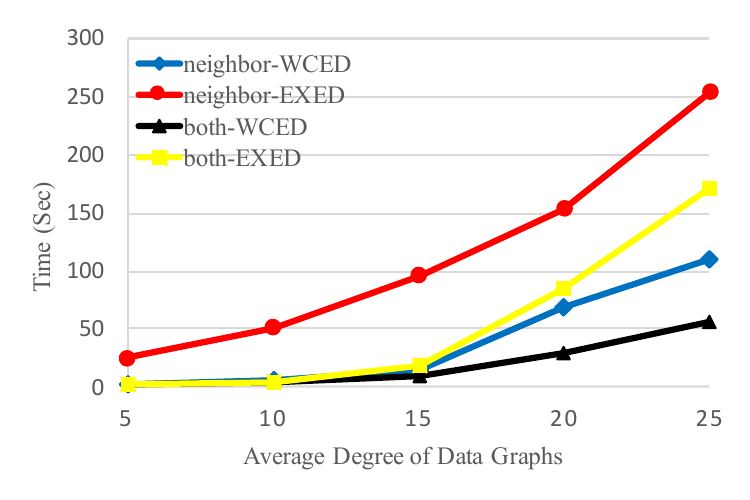}
\caption{Running time varying average degree in data graphs}
\label{fig:density}
\end{minipage}
\hspace*{0.25cm}
\begin{minipage}{.3\textwidth}
\hspace*{-0.5cm}
\includegraphics[width=2.35in]{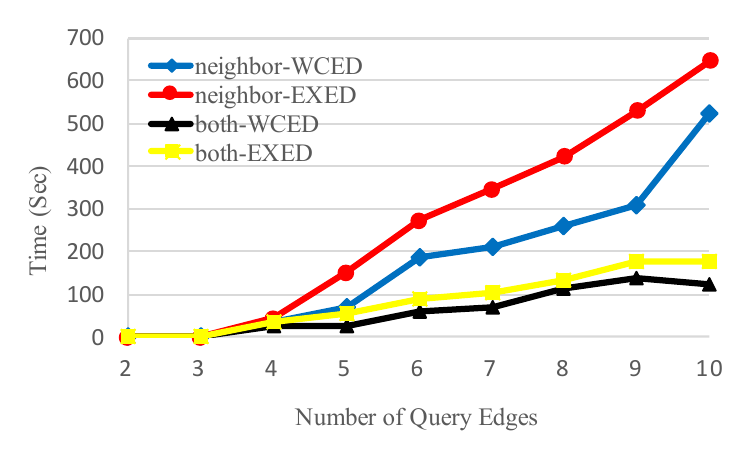}
\caption{Running time varying the number of query edges}
\label{fig:edgenumber}
\end{minipage}
\end{figure*}

\para{Varying the edit distance threshold.}
We varied the edit distance threshold $t$ from $1$ to $5$ and fixed the number of query edges at $8$. Figure~\ref{fig:10ktime} shows that ``neighbor-WCED'' outperforms ``neighbor-EXED'' by a factor of $1.5$ when $t = 5$, reducing the search time more than half in this particular experiment. Comparing ``neighbor-WCED'' and ``both-WCED'', we find that even though there is no clear speedup for adding path-based filtering on top of the neighborhood-based pruning scheme at small thresholds ($t \leq 2$), the performance gap becomes wider at larger thresholds with $\sim 200$ seconds saved when $t=5$. This meets our expectation, because the benefits of using both schemes over ``neighbor'' in pruning power becomes clear when the edit distance threshold increases (Figure~\ref{fig:threshold}).

\para{Varying average degree of data graph.}
In another experiment, we varied the average degree of a node from $5$ to $25$. Figure~\ref{fig:density} shows that ``both-WCED'' has a greater advantage in a data graph with larger average degrees, outperforming ``neighbor-WCED'' and ``neighbor-EXED''. This is because the cost of verifying each candidate depends on the average degree of the data graph, and a larger average degree results in a higher cost of verifying each candidate and a wider gap between ``both-WCED'' and the others.

\para{Varying the number of query edges.}
In another experiment, we varied the number of query edges from $2$ to $10$ with the edit distance threshold fixed at $1$. Figure~\ref{fig:edgenumber} shows that the gap between ``neighbor-WCED'' and ``both-WCED'' (and similarly between ``neighbor-EXED'' and ``both-EXED'') widens, as we increase the number of edges. This is because of the cost of verifying each candidate, which grows exponentially with the number of edges.


Our experiments reveal that adding path-based filtering improves the performance of both EXED and WCED under one or more of these conditions: (1) nodes in data graph have large degrees; (2) $t \geq 2$; (3) the query has over $5$ edges.

%% file: sections/07_related.tex
\section{Other Related Work}
\label{sec:related}

Our method extends exact techniques for graph search by example in order to allow for more flexibility, which is fundamental for exploratory search~\cite{lissandrini2018data}.
Hence, it relates to approaches for query reformulation techniques and approximate query answering.
Compared to existing approximate search methods for graphs, our method (1) allows for matching approximate structures but (2) guarantees the completeness of the result-set without compromising on scalability, and (3) it is the first to consider both edge removal and edge renaming in the edit distance measure.

\para{Graph search.}
Graph search has attracted considerable attention both for the case where the query should retrieve matches from \emph{a large set of distinct graphs} (called a graph database)~\cite{Han:2013:TIT:2463676.2465300} as well as for the case where the search space \emph{is a single large graph}~\cite{mottin2016exemplar,khan2011neighborhood}.
\emph{This work focuses on the latter, and tackles approximate graph search in large graphs.}
In particular, compared to exact Exemplar Queries~\cite{mottin2016exemplar}, which can only find answers that are exactly edge-isomorphic to a query, our algorithm can find relevant subgraphs that are edge-preserving isomorphic to the query after some edit operations.
Jayaram et al.~\cite{jayaram2015querying} present GQBE, an algorithm that takes a set of entities (instead of a graph) and finds the best matching subgraph that includes those entities.
The resulting subgraph may be used as an exemplar query.
GQBE is orthogonal to our work and may be combined with ETEQ, for more efficient query formulations.

\para{Query reformulation.}
A different line of work focuses instead on query reformulation, both for generic graphs~\cite{Vachery:2019:RAQ} and for SPARQL queries~\cite{zheng2016semantic,elbassuoni2011query,huang2012approximating}.
\emph{These works focus on the problem of query rewriting and do not address the problem of fast answer search.}
For the case of SPARQL, queries are represented as ``basic graph patterns'' where some nodes are replaced by variables and an answer should match the structure imposed by the predicates on the edges.
Moreover, they require a corpus of external knowledge to guide the system in deciding what reformulations are allowed.
For instance, the {\it semantic graph edit distance} operation of Zheng et al.~\cite{zheng2016semantic} computes high-level type hierarchies over fact and query graphs (e.g., ``Canada'' may be represented by ``Country''),
while Huang et al.~\cite{huang2012approximating}  require a list of edge preferences and focus on queries that retrieve a specific node.
This requires the user or the system to provide a large amount of task-specific ground truth information, while our algorithms does not require any external information.

For generic graphs, RAQ~\cite{Vachery:2019:RAQ} allows matches that ignore edge labels and compare instead node attributes.
In particular, RAQ targets nodes associated with continuous values to measure similarity and disregards edge labels, yet it does not explicitly address the case of matching different structures.
Though, this type of information and similarity can also be used in our approach in the computation of edit-distance costs.

\para{Approximate queries on graphs.}
To allow for approximate query matching is useful in multiple domains.
In graph databases, this will allow, for instance, the retrieval of molecules with similar compositions (e.g., similar DNA sequences) or 3D objects with similar structures.
A number of works have tackled this problem~\cite{tian2008tale,mongiovi2010sigma,wang2012efficiently,zeng2009comparing,CSIGED}.
Among those,  Wang et al.~\cite{wang2012efficiently} propose  an efficient index for sparse data graphs.
They decompose graphs to small grams (organized by $\kappa$-Adjacent Tree patterns) and use these tree patterns to estimate a lower bound of their edit distance for candidate filtering.
Zeng et al.~\cite{zeng2009comparing} propose a method to compute the edit distance by transforming a graph to a multi-set of star structures and using a path-based index for candidates filtering.
TALE~\cite{tian2008tale} introduces a neighborhood based index (NH-Index) that matches important vertices of a query graph first before extending the match progressively.
SIGMA~\cite{mongiovi2010sigma} is a set-cover based inexact subgraph matching technique.
CSI GED~\cite{CSIGED} focuses on speeding up computation for very large query graphs.
\emph{All the aforementioned approaches work only with graph databases where the number of possible answers is at most the number of graphs in the database}, and their efficient indexes and pre-computations help reduce drastically the search space.

\emph{We focus instead on the case of approximate search over a single large graph.}
In our case, and opposite to the existing literature, the approximation refers only to the structure that is matched and not to the completeness of the answer set, while existing works do not guarantee to return the complete list of all matching answers.

NeMa~\cite{khan2013nema} introduces a similarity measure preserving the proximity of node pairs and label information.
Based on this similarity measure, the authors propose a heuristic for the problem of minimum cost subgraph matching, avoiding the costly subgraph isomorphism and edit distance computation.
This approach focuses on exact matches of node-labels and on the distance between nodes.
Yet, this has been proven to be both less efficient/scalable than exemplar queries and to retrieve less relevant structures~\cite{mottin2016exemplar}.
Dutta et al.~\cite{dutta2017neighbor} study subgraph similarity based on statistical significance, but their notion of statistical significance does not provide a control on the number and types of edits that are allowed in the answer graphs.
Moreover, none of the algorithms above is exact, i.e., they can miss qualifying answers.

The most similar work to ours are SAPPER~\cite{zhang2010sapper} and Semantic Guided Search~\cite{wang2020semantic}.
SAPPER only supports edge removal and does not allow edge renaming, i.e., to match queries with the same structure but different edge labels.
In contrast, ETEQ allows for edge renaming as well.
Moreover, our experiments demonstrate better scalability and response time since SAPPER requires pre-generated random spanning trees that are more costly.
Semantic Guided Search~\cite{wang2020semantic}, instead, assumes queries that define a specific node, called target node.
This imposes a different semantics over queries, since it moves the focus from structures to specific nodes. 
Moreover, it requires for node types and labels to be matched as well.
Also, their focus is on graph search through keyword similarity search, and for this reason they assume that a graph embedding model is also provided in order to measure graph similarity.
Finally, their structural similarity is based on graph weight, where an edge could be replaced with a long path with the same cost, therefore it has no direct control on the amount of structural differences that the answers can contain.


%% file: sections/08_conclusions.tex
\section{Conclusions}
\label{sec:conclusions}
This paper studies the problem of error-tolerant exemplar queries on knowledge graphs. Unlike previous work that supports only exact matching of the labels, the algorithms we developed in this paper allow errors in  both the query and the data graphs. 
We propose two filtering techniques, i.e., neighborhood-based pruning and path-based filtering, and two algorithms, i.e., EXED and WCED, in order to efficiently support the ETEQ queries. 
Moreover, we develop cost models for these algoritmhs, and we estimate and analyze their costs. 
Through a comprehensive experimental evaluation that employs real data, we demonstrate that our algorithms are both efficient and effective, outperforming existing approaches. 

With regards to future work, one direction is to study early termination strategies in the context of Top-k ETEQ queries. Another direction is to study the relationships between different edit operations (e.g., correlations between labels), as well as more adaptive approaches for efficient retrievals.    

%% file: sections/09_acks.tex
\section*{Acknowledgments}
This research is supported by the Natural Sciences and Engineering
Research Council of Canada.

%% file: sections/99_appendix.tex
\newpage
\appendix
\subsection{Proof of Lemmas}
\label{sec:proofs}
%
\noindent\textit{Lemma~\ref{lemma:pl1-k}}
\begin{proof}
This lemma can be proved using the probability subtraction rule:
\begin{align}
&P_D(l_1, l_2, ..., l_k) = \nonumber \\
&P_D(l_2, ...,l_k) - P_D(\neg l_1, l_2, ..., l_k)= \nonumber \\
&P_D(l_2, ...,l_k) - (P_D(\neg l_1, l_3, \ldots, l_k) - P_D(\neg l_1, \neg l_2, ..., l_k))= \nonumber \\
&P_D(l_2, l_3, ...,l_k) - P_D(\neg l_1, l_3, \ldots, l_k) + P_D(\neg l_1, \neg l_2, l_4, \ldots, l_k) \nonumber \\
&- P_D(\neg l_1, \neg l_2, \neg l_3, ..., l_k)= \nonumber\\
&...\nonumber\\
&\sum_{i=2}^k (-1)^{i-1}P_D(\neg l_j,\ldots,\neg l_{i-1}, l_{i+1}, \ldots, l_k)\nonumber\\
&+ (-1)^k P_D(\neg l_1, \neg l_2, ..., \neg l_k) + P_D(l_2, l_3, ...,l_k).
\label{eq:P_Dderive}
\end{align}

If we expand $P_D(l_2,\ldots,l_k)$ further using the equation above, we will have a set of terms that look similar to the first and the second terms in Eq~\ref{eq:P_Dderive} and the base case $P_D(l_k)$. For the base case, we have $P_D(l_k)=(1-(1-Sel(l_k))^D$. We also know that $P_D(\neg l_j, \ldots, \neg l_k)=(1-\sum_{i=j}^k Sel(l_i))^D$ assuming independence. Putting these pieces together will give the statement of the lemma.

\end{proof}

\noindent\textit{Lemma~\ref{lemma:comp}}
\begin{proof}

Using Equations~\ref{eq:cost-verifying-n}, the cost of verifying wildcard queries for WCED with edit distance $1$ can be written as
\begin{align*}
Cost_{wc} =\sum_{k = 1}^{|E_q|}\sum_{i=1}^{|E_q|}\prod_{j = 1}^i\hat{D}*Sel(l_{k,j})
\end{align*}

Let $l_1,\ldots,l_k$ denote the labels in an increasing order of selectivity. Since the edges in a query are verified in an increasing order of their selectivity, for those wildcard queries where $l_j (1\leq j \leq i)$ is not set to the wildcard, the edge with label $l_i$ is verified at the $i^{th}$ step of the simulation, and the cost of verifying the edge is $\hat{D}^i\prod_{j=1}^iSel(l_j)$. There are $|E_q| - i$ such wildcard queries. For those wildcard queries where $l_j (1\leq j \leq i)$ is set to the wildcard, the edge with label $l_{i+1}$ is verified at the $i^{th}$ step of the simulation, and the cost of verifying this edge is $\hat{D}^i\prod_{j=1}^{i+1}Sel(l_j)$, where $l_j \not= l_m$.

Let $T_i$ be
\begin{equation}
\label{eq:tl}
T_i = \sum_{k=1}^{i} \prod_{m=1}^{i} Sel(l_{k,m}) \text{ where } l_{k,m} \not= l_k.
\end{equation}

The sum of verification costs for those wildcard queries with $l_m (1\leq m \leq i - 1)$ set to the wildcard at the $i^{th}$ step of the simulation is $\hat{D}^i(T_{i+1} - \hat{D}^i\prod_{j=1}^{i-1}Sel(l_j))$.
Then, the sum of verification costs for WCED can be written as
\begin{align*}
Cost_{wc}&=\sum_{i=1}^{|E_q| - 1}\hat{D}^i((|E_q| - i - 1)\prod_{j=1}^i Sel(l_j) + T_{i+1})\\
&+\hat{D}^{|E_q|}T_{|E_q|}.
\end{align*}

Using Equations~\ref{eq:st-cases} and~\ref{eq:ed-verify}, the verification cost of EXED with edit distance threshold $1$ can be written as
\begin{align*}
Cost_{ex}&=\hat{D} + \sum_{i=2}^{|E_q|}(\hat{D}^{i}\sum_{k=1}^{i-1} (1 - Sel(l_k))\prod_{j=1}^{i}Sel(l_{k,j})\\
&+ \hat{D}^{i}\prod_{j=1}^{i-1}Sel(l_j)), \text{where } l_{k,j} \not= l_k.
\end{align*}
By replacing the terms in above equation with Equation~\ref{eq:tl}, $Cost_{ex}$ can be written as
\[
Cost_{ex}=\hat{D} + \sum_{i=2}^{|E_q|}\hat{D}^{i}(T_{i} - (i-1)\prod_{j=1}^{i}Sel(l_j)).
\]
Then, the difference between two costs $\Delta_{cost}$ can be written as
\begin{align*}
&\Delta_{cost} = \hat{D}(1 - (|E_q| - 1)Sel(l_1) - Sel(l_2)) \\
&+ \sum_{i=2}^{|E_q| - 1}\hat{D}^i(T_i - T_{i+1} -  (|E_q| - 2)\prod_{j=1}^i Sel(l_j))\\
&- (|E_q| - 1)\hat{D}^{|E_q|}\prod_{i=1}^{|E_q|}Sel(l_i).
\end{align*}
Since query edges are visited in increasing order of label selectivities, we have an inequality as follows
\[
Sel(l_1) \leq Sel(l_i) \leq 1.
\]
With the inequality above, we have
\[
iSel(l_1)^{i-1}\leq T_i \leq i
\]
Using the both inequalities above, the upper bound of $\Delta_{cost}$ can be written as
\begin{align*}
&\Delta_{cost} \leq \hat{D}(1 - {|E_q|}Sel(l_1)) - (|E_q| - 1)\hat{D}^{|E_q|}Sel(l_1)^{|E_q|}\\
&+ \sum_{i=2}^{|E_q| - 1}\hat{D}^i(i -(|E_q| + i - 1) Sel(l_1)^i).
\end{align*}
Let $F_n(x)$ denote the upper bound of $\Delta_{cost}$ using $x$ to denote $Sel(l_1)$ and $n$ to denote the number of query edges. To show the correctness of the Lemma~\ref{lemma:comp}, we prove that $F_{|E_q|}(x) \leq 0$ with different number of edges when the conditions in the Lemma holds using mathematical induction.
\basis
$n = 2$: $F_2(x)$ can be written as
\[
F_2(x) = \hat{D}(1 - 2x) - \hat{D}^2x^2
\]
When $x = \frac{1}{\sqrt{\hat{D}}}$, we have $F_2(x)$.
\[
\hat{D}(1 - 2x) - \hat{D}^2(\frac{1}{\sqrt{\hat{D}}})^2 = \hat{D}(-2x)<0.
\]
We also know that the derivative of $F_2(x)$ is
\[
\frac{\partial F_2}{\partial x} = -2\hat{D} - 2\hat{D}^2x < 0.
\]
Combining two facts above, we know that $F_2(x) < 0$ when $Sel(l_1) > \frac{1}{\sqrt{\hat{D}}}$.
\ih
the Lemma holds when the query has $k$ edges.
\begin{align*}
F_k(x) &= \hat{D}(1 - kx) - (k - 1)\hat{D}^kx^k\\
&+ \sum_{i=2}^{k - 1}\hat{D}^i(i -(k + i - 1) x^i) < 0\\
& \text{subject to }  x > \frac{1}{\sqrt[k]{\hat{D}}}.
\end{align*}
Note that the derivative of  $F_k(x)$ is
\[
\frac{\partial F_{k}}{\partial x}= -k - k(k-1)\hat{D}^kx^{k-1}-\sum_{i=2}^ki(k+i-1)x^{i-1} < 0.
\]
\is Using $F_k(x)$ to substitute some terms in $F_{k+1}(x)$,  $F_{k+1}(x)$ can be written as
\begin{align*}
&F_{k+1}(x) = \hat{D}(1 - (k+1)x) - k\hat{D}^{k+1}x^{k+1}\\
&+ \sum_{i=2}^{k}\hat{D}^i(i -(k + i ) x^i) =\\
 &F_k(x) - \hat{D}x -  \sum_{i=2}^{k-1}x^i + D^k(k - (k+1)x^k) + k\hat{D}^{k+1}x^{k+1}.
\end{align*}
When  $x = \frac{1}{\sqrt[k+1]{\hat{D}}}$, after replacing the $x$ with the value in the last term and combining the last two terms, $F_{k+1}(x)$ can be written as
\begin{align*}
&F_{k+1}(\frac{1}{\sqrt[k+1]{\hat{D}}}) = F_k(x) - \hat{D}x -  \sum_{i=2}^{k-1}x^i + D^k(k - (k+1)x^k) \\
&+ k\hat{D}^{k+1}(\frac{1}{\sqrt[k+1]{\hat{D}}})^{k+1} = F_k(x) - \hat{D}x -  \sum_{i=2}^{k-1}x^i -(k+1)D^kx^k
\end{align*}
Since $x = \frac{1}{\sqrt[k+1]{\hat{D}}} > \frac{1}{\sqrt[k]{\hat{D}}}$, $F_k(x) < 0$ and the rest of terms are also negative, we have
\[
F_{k+1}(\frac{1}{\sqrt[k+1]{\hat{D}}}) < 0.
\]
We also know that the derivative of $F_{k+1}(x)$ is negative.
\[
\frac{\partial F_{k+1}}{\partial x} = \frac{\partial F_{k}}{\partial x} - \hat{D} - \sum_{i=2}^{k-1}ix^{i-1} - k(k+1)\hat{D}^kx^{k-1} < 0.
\]
Combining two facts above, we know that $F_{k+1}(x) < 0$ when $Sel(l_1) > \frac{1}{\sqrt[k+1]{\hat{D}}}$.
\end{proof}
\subsection{Query Set}
\label{sec:query_set}
The AOL query set used in our comparison with exemplar queries in Section ~\ref{sec:comp}, formatted as``$<$subject$>$ $<$predicate$>$ $<$object$>$'', are as follows:
\begin{enumerate}
\item D influenced Swift;\\
Scala influenced Swift;\\
Ruby influenced Swift;\\
Rust influenced Swift;\\
Swift languages Function programming;\\
Swift languages Procedural programming;\\
Swift languages Generic programming;\\
Swift developer Treehouse.
\item Myocardial infarction people Anatole Dauman;\\
Anxiety symptom\_of depression;\\
Stress diseases Myocardial infarction;\\
Stress associated\_land\_cover\_s Graves' disease;\\
Peptic ulcer risk\_factors Stress;\\
Graves' disease symptoms Anxiety;\\
Stress diseases Conversion disorder.
\item Going Upriver executive\_produced\_by Marc Abrams;\\
The Main Event writer Michael Benson;\\
Sun Ahso Rises commanders Michael Benson;\\
Sun Ahso Rises writer Marc Abrams;\\
Marc Abrams episodes\_written Sun Ahso Rises;\\
Marc Abrams episodes\_written The Main Event.
\item Going Upriver executive\_produced\_by Marc Abrams;\\
The Main Event writer Michael Benson;\\
Sun Ahso Rises commanders Michael Benson;\\
Sun Ahso Rises writer Marc Abrams;\\
Marc Abrams episodes\_written Sun Ahso Rises;\\
Marc Abrams episodes\_written The Main Event.
\item Frederick County contains Ole Orchard Estates;\\
Frederick County events Second Battle of Winchester;\\
Frederick County  buildings\_occupied North Mountain;\\
Frederick County contains Echo Village;\\
Frederick County people\_born\_here James Brenton (1740–1782);\\
Frederick County contains Green Acres;\\
Frederick County contains US Census 2000 Tract 51069050100.
\item Research subject\_of Carnegie Moscow Center;\\
Research works Hot talk, cold science;\\
Research works Person or Persons Unknown;\\
Research address Stanford University School of Medicine;\\
Research schools\_of\_this\_kind Indian Institute of Forest Management;\\
Research organizations\_of\_this\_type Stanford Radiology.
\item Valve Corporation games\_developed Half-Life 2;\\
Valve Corporation games\_published Wolfenstein 3D;\\
Valve Corporation games\_published The Maw;\\
Valve Corporation games\_developed CS Online;\\
Valve Corporation games\_published Half-Life 2;\\
Valve Corporation is\_reviewed Place founded;\\
Valve Corporation games\_published CS Online.
\item Scheme influenced Haskell;\\
Scheme influenced Clojure;\\
Scheme influenced LFE;\\
Scheme influenced Dylan;\\
Scheme influenced\_by Lisp;\\
Scheme parent\_language Lisp.
\item NetBSD supported\_architectures x86;\\
x86 manufacturers United Microelectronics Corporation;\\
NetBSD supported\_architectures ARM architecture;\\
Great Giana Sisters game The Great Giana Sisters;\\
The Great Giana Sisters governing\_body NetBSD;\\
x86 manufacturers Cyrix;\\
NetBSD parent\_os 386BSD.
The Great Giana Sisters platforms Dreamcast
\item Xbox 360 games\_on\_this\_platform Garret the Slug;\\
NBA 2K11 platforms Xbox 360;\\
Xbox 360 games\_on\_this\_platform Deus Ex: Human Revolution;\\
Halo 3 platform Xbox 360;\\
Microsoft Corporation games\_published The Maw;\\
Xbox 360 games\_on\_this\_platform Rainy Woods;\\
Halo 3 publisher Microsoft Corporation;\\
Xbox 360 games\_on\_this\_platform Halo 3.
\end{enumerate}

The QALD4 queries used in our evaluation are as follows:
\begin{enumerate}
\item Which books by Kerouac were published by Viking Press?\\
{\bf e.g.} On\_the\_Road\\
{\bf predicates} ($\ast$Kerouac, notableWork, X), (X, author, $\ast$Kerouac), (X, publisher, Viking\_Press)
\item Which states of Germany are governed by the Social Democratic Party?\\
{\bf e.g.} Berlin\\
{\bf predicates} (X, leaderParty, Social\_Democratic\_Party\_of\_Germany), (Social\_Democratic\_Party\_of\_Germany , headquarter, X), (X, leader, a), (a, party, Social\_Democratic\_Party\_of\_Germany), ($\ast$, state, X)
\item Which television shows were created by Walt Disney?\\
{\bf e.g.} The\_Mickey\_Mouse\_Club\\
{\bf predicates} (X, creator, Walt\_Disney), (X, company, $\ast$Walt\_Disney$\ast$),
(X, format, $\ast$television$\ast$), (X, format, $\ast$show)
\item Which actors were born in Germany?\\
{\bf e.g.} Briana\_Banks\\
{\bf predicates} (X, birthPlace, Germany), (X, ethnicity, German$\ast$), (X, numberOfFilms, $\ast$),
(X, weight, $\ast$), (X, height, $\ast$), ($\ast$, starring, X), (X, birthDate, $\ast$)
\item Give me all people that were born in Vienna and died in Berlin.\\
{\bf e.g.} Hilde\_K\%C3\%B6rber\\
{\bf predicates} (X, birthPlace, Vienna), (X, deathPlace, Berlin)
\item Which companies work in the aerospace industry as well as in medicine?\\
{\bf e.g.} Makino\\
{\bf predicates} (X, industry, Aerospace), (X, industry, Medicine), (X, type, $\ast$company)
\item Which languages are spoken in Estonia?\\
{\bf e.g.} Estonian\_language\\
{\bf predicates} (X, spokenIn, Estonia), (Estonia, officialLanguage, X), (Estonia, language, X)
\item Give me all soccer clubs in Spain.\\
{\bf e.g.} Albacete\_Balompi\%C3\%A9\\
{\bf predicates} (X, ground, Spain), ($\ast$, managerClub, X)
\item Which countries adopted the Euro?\\
{\bf e.g.} Andorra\\
{\bf predicates} (X, currency, Euro), ($\ast$, country, X)
\item In which military conflicts did Lawrence of Arabia participate?\\
{\bf e.g.} Arab\_Revolt\\
{\bf predicates} (T.\_E.\_Lawrence, battle, X), (X, commander, T.\_E.\_Lawrence), (X, isPartOfMilitaryConflict, $\ast$), ($\ast$, isPartOfMilitaryConflict, X)
\item Give me the capitals of all countries in Africa.\\
{\bf e.g.} Luanda\\
{\bf predicates} ($\ast$, capital, X), ($\ast$Africa$\ast$, location, X), ($\ast$Africa$\ast$, city, X)
\item Give me all islands that belong to Japan.\\
{\bf e.g.} Kyushu\\
{\bf predicates} (X, country, Japan), ($\ast$Japan$\ast$, place, X), ($\ast$X$\ast$, location, Japan)
\item Which airports are located in California, USA?\\
{\bf e.g.} Moffett\_Federal\_Airfield\\
{\bf predicates} (X, location, California), (X, iataLocationIdentifier, $\ast$), (X, icaoLocationIdentifier, $\ast$), (X, faaLocationIdentifier, $\ast$)
\item Which Chess players died in the same place they were born in?\\
{\bf e.g.} Paul\_Morphy\\
{\bf predicates} ($\ast$Chess$\ast$, editor, X), (X, worldChampionTitleYear, $\ast$), (X, birthPlace, a), (X, deathPlace, a)
\item Which capitals in Europe were host cities of the summer olympic games?\\
{\bf e.g.} Amsterdam\\
{\bf predicates} (a, capital, X), ($\ast$Europe$\ast$, location, a), ($\ast$, city, X)
\item Give me all cars that are produced in Germany.\\
{\bf e.g.} Porsche\_928\\
{\bf predicates} (X, assembly, Germany), (X, manufacturer, $\ast$), (X, productionStartYear, $\ast$), (X, productionEndYear, $\ast$), (X, class, $\ast$), (X, bodyStyle, $\ast$), (X, engine, $\ast$), (X, transmission, $\ast$), (X, wheelbase, $\ast$), (X, length, $\ast$), (X, width, $\ast$), (X, weight, $\ast$)
\item Give me all actors starring in movies directed by William Shatner.\\
{\bf e.g.} Leonard\_Nimoy\\
{\bf predicates} (a, starring, X), (a, director, William\_Shatner)
\item Give me all actors starring in Last Action Hero.\\
{\bf e.g.} Arnold\_Schwarzenegger\\
{\bf predicates} (Last\_Action\_Hero, producer, X), (Last\_Action\_Hero, starring, X)
\item Give me all video games published by Mean Hamster Software.\\
{\bf e.g.} Myst\\
{\bf predicates} (Mean\_Hamster\_Software, product, X), (X, publisher, Mean\_Hamster\_Software), (X, genre, $\ast$game)
\item Who produced films starring Natalie Portman?\\
{\bf e.g.} Patrice\_Ledoux\\
{\bf predicates} (a, producer, X), (a, starring, Natalie\_Portman)
\end{enumerate}